\documentclass[useAMS,usenatbib]{mn2e}
\usepackage{graphicx,amssymb}

\usepackage[T1]{fontenc}
\usepackage{ae,aecompl}

\title[Cosmic ray acceleration and escape from supernova remnants]
{Cosmic ray acceleration and escape from supernova remnants}
\author[A.R. Bell, K.M. Schure, B. Reville and G. Giacinti]
{A. R. Bell\thanks{E-mail:t.bell1@physics.ox.ac.uk},
K.M. Schure, B. Reville and G. Giacinti\\
Clarendon Laboratory, University of Oxford, Parks Road, Oxford OX1 3PU, UK\\ }

\begin{document}
\date{}
\pagerange{\pageref{firstpage}--\pageref{lastpage}} \pubyear{2013}
\maketitle
\label{firstpage}
\begin{abstract}
Galactic cosmic ray (CR) acceleration to the knee in the spectrum at a few PeV is only possible if 
the magnetic field ahead of a supernova remnant (SNR) shock is strongly amplified by CR escaping the SNR.
A model formulated in terms of the electric charge carried by escaping CR predicts the maximum CR energy
and the energy spectrum of CR released into the surrounding medium.
We find that historical SNR such as Cas A, Tycho and Kepler may be expanding too slowly
to accelerate CR to the knee at the present time.
\end{abstract}

\section{Introduction}

During diffusive shock acceleration (DSA) cosmic rays (CR) gain energy by repeatedly passing back and forth 
between the upstream and downstream plasmas 
(Krymskii 1977, Axford et al 1977, Bell 1978, Blandford \& Ostriker 1978).
CR diffuse ahead of the shock to form a precursor with an exponential scaleheight $D_u/u_s$
where $u_s$ is the shock velocity and $D_u$ is the CR diffusion coefficient upstream of the shock.
The average dwell-time spent upstream of the shock between 
successive shock crossings is $4D_u/cu_s$ for relativistic particles (Bell 2012).
This, along with the corresponding downstream dwell-time, determines the rate at which CR 
are accelerated.
Lagage \& Cesarsky (1983a,b) showed that the time taken for CR acceleration is 
$t_{accel}=4D_u/u_s^2+4D_d/u_d^2$ where $D_d$ is the downstream diffusion coefficient
and $u_d$ is the downstream fluid velocity in the shock rest frame.
Since $u_d=u_s/4$ for a strong shock it might appear that the downstream dwell-time determines the
acceleration rate, but we can expect $D_d \ll D_u$ partly because the magnetic field is increased by compression by the shock, 
partly because the compressed downstream magnetic field is more closely perpendicular to the shock normal,
and partly because the downstream field is disturbed and more irregular after passing through the shock.
If the downstream and upstream dwell-times are the same
$t_{accel}=8D_u/u_s^2$.
A further common assumption is that Bohm diffusion applies:  $D_u=cr_g$ or $D_u=cr_g/3$ where $r_g$ is the
CR Larmor radius.  
This assumes a diffusion model in which CR are scattered by irregularities in the magnetic field
such that the scattering mean free path is of the order of the CR Larmor radius.
There is some observational evidence for Bohm diffusion (Stage et al 2006, Uchiyama et al 2007). 
Furthermore if the mean free path were much larger than a Larmor radius acceleration by SNR
to the knee in the Galactic CR spectrum would
be very difficult. 
The maximum CR energy is determined by the condition
that $t_{accel}$ cannot exceed the age $t_{SNR}$ of an SNR (Lagage \& Cesarsky 1983a,b). 
Assuming $D_u=cr_g$,
the maximum CR energy $T_{max}$ in eV is given by
$T_{max}=0.03B_{\mu G}u_7^2 t_{1000}$PeV
where $B_{\mu G}$ is the upstream magnetic field in microGauss,
$u_7$ is the shock velocity in units of 10,000 km s$^{-1}$ 
and $t_{1000}$ is the SNR age in 1000's of years.
For a typical young SNR expanding into the interstellar medium
without magnetic field amplification
$B_{\mu G}=3$, $u_7=0.5$ and $t_{1000}=0.4$, giving $T_{max}=0.01$PeV
which falls a factor of $\sim 100$ short of that required to explain the Galactic CR spectrum.
SNR in the Sedov phase do not fare better.  Their shock velocities decrease
in proportion to ${\rm time}^{-3/5}$ and ${\rm radius}^{-3/2}$,
so little benefit accrues from their larger age and radius.
This posed a serious problem for diffusive shock acceleration as
an explanation of the Galactic spectrum until it was shown 
that a plasma instability driven by streaming CR in the upstream precursor could amplify the 
magnetic field ahead of the shock and facilitate rapid acceleration to higher energies
(Lucek \& Bell 2000, Bell 2004, 2005).

The phenomenon of magnetic field amplification provides a
mechanism by which the CR energy can be raised significantly beyond 0.01PeV,
but there remains the question of why the fields are amplified to the observed magnitude,
up to 100's $\mu{\rm G}$ in the historical SNR 
(Vink \& Laming 2003, Berezhko et al 2003, V\"{o}lk et al 2005),
and why CR are accelerated to a few PeV rather than 0.1 or 10 PeV.
We try to answer these questions by examining the self-consistent interaction 
between streaming CR, behaving kinetically, 
and the upstream plasma behaving magnetohydrodynamically.

It has been known for many years that the escape of CR upstream of the shock is an important
part of the overall acceleration process as discussed below in the final paragraphs of section 3.
We find that the combined CR-MHD system organises itself
to allow a suitable number of CR to
escape upstream. 
The CR drive magnetic field amplification which in turn
regulates the number of escaping CR.
If a smaller number of CR escaped, the magnetic field would be insufficiently amplified
to confine and accelerate the CR.
If a larger number of CR escaped, the magnetic field would grow too rapidly to allow their escape.
Hence a self-regulating system is set up that determines the number and maximum energy of escaping CR.

The paper is organised as follows.  Sections 2 and 3 present 
approximate calculations showing how a limit on the CR energy
is placed by the need for CR to drive 
magnetic field amplification by escaping upstream.
Sections 4 to 7 describe Vlasov-Fokker-Planck (VFP) simulations that support
the arguments of sections 2 and 3.
Sections 8 to 11 apply the results to supernova remnants 
and the Galactic cosmic ray spectrum.
Readers unfamiliar with VFP simulations may wish to
read sections 1 to 3 and 8 to 11 before returning to the computational
validation and illustration in sections 4 to 7.


\section{Conditions for strong magnetic field amplification}

We assume that magnetic field is generated by the non-resonant hybrid (NRH) instability described by Bell (2004).
This is one of a class of plasma instabilities driven by CR streaming.
In its simplest form, CR have a Larmor radius much greater than the wavelength of spiral perturbations in 
a zeroth order uniform magnetic field.
Because of their large Larmor radius the streaming CR, carrying an electric current density ${\bf j}_{CR}$, 
are essentially undeflected by the
perturbed field but the ${\bf j}_{CR}\times {\bf B}$ force acts towards 
the centre of the spiral.
A corresponding reactive force acts on the background plasma
to expand the spiral.
This stretches and increases the magnitude of the perturbed magnetic field, thereby increasing the
${\bf j}_{CR}\times {\bf B}$ force in a positive feedback loop that drives the instability.
NRH appears to be the most rapidly growing instability driven by CR streaming on a relevant scalelength.
Other instabilities that have attracted significant attention are the resonant Alfven instability 
(Kulsrud \& Pearce 1969, Wentzel 1974) 
that grows with spatial wavelengths spatially resonant with the CR Larmor radius 
and the Weibel instability (Weibel 1959) that grows 
quickly on the spatial scale of an electron or proton collisionless skin depth, 
$c/\omega _{pe} = 5.3 (n_e/{\rm cm}^{-3})^{-1/2}$km, $c/\omega _{pi}=\sqrt{m_p/m_e}c/\omega _{pe}$.
The Weibel instability is important for interactions engaging thermal or mildly suprathermal electrons and ions,
and may be effective for CR acceleration to low energies, 
but it grows on a scale too small to scatter PeV ions which have a Larmor radius of $\sim 10^9 c/\omega _{pe}$.
The Alfven instability grows on the desired spatial scale but grows less quickly 
than the NRH instability in the SNR shock environment.
Instabilities that grow on scales larger than the CR Larmor radius 
(Bykov et al 2011, Drury \& Falle 1986, Drury \& Downes 2012, Malkov \& Diamond 2009,  
Rogachevskii et al 2012, Schure \& Bell 2011)
tend to have longer
growth times but turbulent amplification may 
increase the growth rate (Bykov et al 2011).
An extended discussion of instabilities driven by cosmic ray streaming
can be found in Schure et al (2012).

From the Lagage \& Cesarsky (1983a,b) limit as described in section 1
it is clear that proton acceleration to a few PeV is only possible if the magnetic field
is strongly amplified above its characteristic interstellar value of a few ${\mu}G$.
The following argument places an upper limit on
the maximum energy of a CR that can be strongly scattered by 
a magnetic field growing on the scale of a CR Larmor radius.
The background thermal plasma is highly magnetised on the scale of a PeV CR Larmor radius.
Consequently the magnetic field is `frozen in' to the thermal plasma.
Magnetic field amplification occurs
as the plasma moves and stretches field lines.
Assuming the perturbed magnetic field does not far exceed the initial field, the
${\bf j}_{CR}\times {\bf B}$ force displaces a plasma element a maximum distance
$s_{max}\sim (j_{CR} Bt^2)/ 2\rho $ in time $t$
where $B$ is the initial seed field.  
For the stretched magnetic field
to strongly scatter CR as required for Bohm diffusion
the displacement must grow to the order of a CR Larmor radius,
that is $s_{max} \sim T/cB$ where $T$ is the CR energy in eV.
The CR current density ${\bf j}_{CR}$ carries an energy flux ${\bf j}_{CR}T$
in the upstream plasma rest frame
which cannot far exceed the energy flux $\rho u_s^3/2$ carried by the upstream plasma into the shock,
where $\rho$ is the upstream mass density.
We define a CR acceleration efficiency $\eta$ such that $j_{CR}T=\eta \rho u_s^3$.
We then have two equations: $s_{max} = (j_{CR}Bt^2)/ 2\rho \sim T/cB $
and $j_{CR} = \eta \rho u_s^3/T$.
When combined, they yield a CR energy $T \sim (\eta c u_s^3)^{1/2}Bt$.
This expression for $T$ is equivalent to 
$T \sim 1.5 (\eta u_7^3)^{1/2}B_{\mu {\rm G}} t_{1000}$PeV
where  $u_7$ is the shock velocity in units of $10,000{\rm km\ s}^{-1}$,
$B_{\mu {\rm G}}$ 
is the seed magnetic field in $\mu {\rm G}$ from which amplification begins,
and $t_{1000}$ is the time in 1000's of years.
Characteristically for young SNR, $u_7=0.5$,  $t_{1000}=0.4$,
$\eta=0.03$ (see Appendix A) and  $B_{\mu {\rm G}}=3 $ where the magnetic field 
of a few $\mu {\rm G}$ represents the seed field from which the instability grows.
With these estimates, $T\sim 0.1$PeV, which is an order of magnitude less than the
energy of the knee.
This estimate is independent of any detailed instability theory except for the assumption that 
magnetic field amplification takes place through plasma motions generated
by the ${\bf j}_{CR}\times {\bf B}$ force
acting on the scale of a CR Larmor radius.
It highlights the difficulty of amplifying magnetic field by 
orders of magnitude on the scale of the Larmor radius of a PeV proton
and the need for an instability that can provide non-linear growth of the magnetic field.

The NRH instability offers a way round this difficulty by initially growing very rapidly
on a scale much less than the CR Larmor radius.  
Since the small-scale magnetic field grows unstably
the ${\bf j}_{CR}\times {\bf B}$ force grows exponentially in time.
The scale-size increases to the CR Larmor radius during non-linear growth.
The NRH growth rate is $\gamma = (kj_{CR}B_0/\rho)^{1/2}$ where $B_0$ is the zeroth order magnetic field
and $k$ is the wavenumber on which the instability grows.  
The maximum NRH growth rate is
$\gamma _{max}=0.5 j_{CR}(\mu _0/\rho)^{1/2}$ which occurs at a wavenumber
$k_{max}=0.5 \mu _0 j_{CR}/B_0$.
$\gamma _{max}$ is $(k_{max}r_g)^{1/2}$ times larger than the NRH growth rate 
for $kr_g=1$,
thus allowing the magnetic field and the ${\bf j}_{CR}\times {\bf B}$ force
to increase rapidly.
The NRH and Alfven growth rates are similar at $kr_g=1$ (see Appendix E and
Bell 2004).
Using the above nomenclature,
$k_{max}r_g=\eta u_s^3/cv_A^2= 2\times 10^4 \eta _{0.03} u_7^3n_e B_{\mu {\rm G}}^{-2}$
where $v_A$ is the Alfven speed, $\eta =0.03 \eta_{0.03}$ and $n_e$ is the electron density
in cm$^{-3}$.
For a discussion of the value of $\eta$ see Appendix A and Bell (2004).
The growth time of the fastest growing mode is then
$\gamma_{max}^{-1}=50 \eta_{0.03^{-1}}n_e^{-1/2}u_7^{-3} T_{PeV}$ years
where $T_{PeV}$ is the energy in PeV of the CR driving the instability.
 $\gamma_{max}^{-1}=400$ years for 
 $\eta_{0.03}=1$, $n_e=1$, $u_7=0.5$
 which implies that even the NRH instability struggles to amplify the magnetic field
 sufficiently to accelerate CR to PeV energies in the historical SNR.
 For growth by many e-folding the NRH instability must be driven by CR with energies less than 1PeV.

\begin{figure*}
\includegraphics[angle=270,width=12cm]{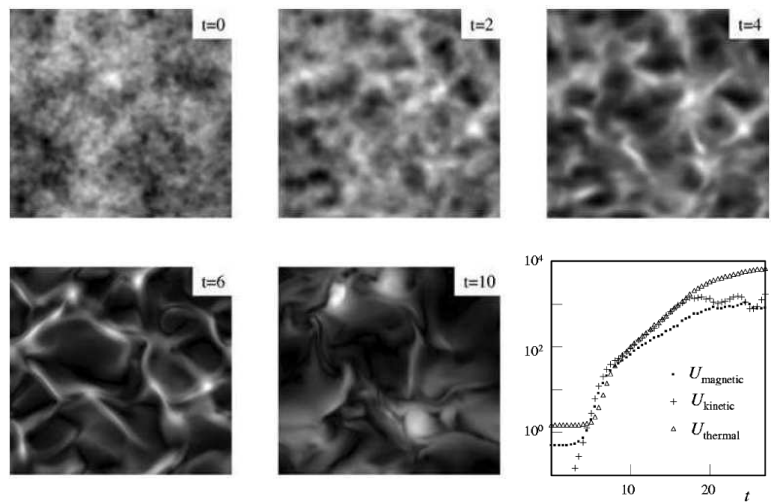}
\caption{Plots of the magnitude of a magnetic field driven by CR streaming into the page
(2D slices of a 3D simulation).
The field grows from noise on a small spatial scale at $t=0$.
By $t=10$ the spatial scale has grown to the size of the computational box.
The graph shows how the mean magnetic, kinetic and thermal energy densities increase with time.
This figure is composed from figures 3 and 4 of Bell (2004) where further details can be found.
The units of time are such that $\gamma_{max}=1.26$.
}
\label{fig:figure1}
\end{figure*}

As remarked above,
the fastest growing mode grows on a scale $k_{max}^{-1}$ whereas efficient CR scattering
requires fluctuations in the magnetic field on a scale $r_g$.
The above analysis showed that
$k_{max}r_g = 2\times 10^4 \eta _{0.03} u_7^3n_e B_{\mu {\rm G}}^{-2}$.
The instability initially grows on a scale too small to effectively scatter PeV CR.
Two factors save the situation.
Firstly, as the magnetic field grows from a few $\mu$G to a few hundred $\mu$G the CR Larmor radius
 decreases by the corresponding factor of $\sim 100$.
 Secondly, the characteristic scalelength of the structured magnetic field increases during
 non-linear growth of the instability.
 Figure 1 is drawn from figures 3 and 4 of Bell (2004).
 The graph of energy densities shows how the magnitude of the magnetic field grows to many times its seed value
 in a time $\sim 10 \gamma_{max}^{-1}$.
 The characteristic scalelength grows during this time as shown by
 the grey-scale images.
 By the time $t=10$ the scalelength has grown to the size of the computational box from its initially 
much smaller scale.
Both $k$ and $r_g$ decrease by a large factor by the time
$10 \gamma_{max}^{-1}$.  
The simulation reproduced in figure 1 had a fixed CR current so it was impossible to demonstrate 
strong CR scattering by the magnetic field after  $t=10 \gamma_{max}^{-1}$,
but the figure provides strong evidence for 
rapid initial growth of the fastest growing modes into
large scale magnetic structures able to scatter CR on the scale of a Larmor radius.


\section{CR escape upstream of the shock}

The previous section does not provide an estimate of the magnitude of
the amplified magnetic field, but it does define conditions under
which strong magnetic field amplification occurs.
CR acceleration to PeV energies requires strong magnetic field amplification
which fixes the number of instability e-foldings to the range 5-10.
This in turn fixes the CR current.
From figure 1 we take the condition for strong magnetic field amplification in a particular volume of upstream
plasma to be that $\int \gamma_{max}dt \sim 5$ in the time before the shock overtakes it.
Since $\gamma_{max}=0.5 j_{CR} \sqrt{\mu_0/\rho}$, the condition for field amplification is
$$
Q_{CR}=\int j_{CR}dt=10\sqrt{\rho/\mu_0}
\eqno{(1)}
$$
where $Q_{CR}$ is the total electric charge of CR passing through unit surface area upstream of
the shock before the shock arrives.
(It makes no difference whether the CR passing through the plasma
have high or low energies.
It is only the number of CR escaping upstream that counts.)
Since only the highest energy CR escape,
the areal charge $Q_{CR}$ is carried by the highest energy CR being accelerated.
The Larmor radius of a PeV proton in the interstellar medium is 
comparable with the radius of the historical SNR and its scattering mean free path is very much greater than the SNR radius.  
These escaping CR pass into the interstellar medium with low probability of further encounter with the SNR shock.
Strongly scattered CR at lower energies are confined by the shock
and are advected away downstream after acceleration.

Suppose that the CR distribution follows a $p^{-4}$ power law up to a momentum $p_{max}$, then
in steady state the electrical current of CR escaping 
in a small band of energies above the energy
$eT_{max}=cp_{max}$ is
$$
j_{CR} = eu_s \pi p_{max}^3 f_0(p_{max})
\eqno{(2)}
$$
where $f_0$ is the isotropic part of the CR distribution 
in momentum space at the shock.
Equation (2) is derived from equation (11a) below by integrating in space
across the shock and deriving the rate at which CR reach the momentum $p_{max}$.
The derivation can be found in Appendix B.
In comparison the CR pressure at the shock, 
where the CR distribution extends down to 
$p=mc$ and $m$ is the proton mass is
$$
P_{CR}
= \frac {4\pi}{3} cp_{max}^4 f_0(p_{max}) \ln \left( \frac{p_{max}}{mc} \right)
\eqno{(3)}
$$
giving
$$
j_{CR}=0.05 \frac{u_s P_{CR}}{T_{max}}
\eqno{(4)}
$$
where we have assumed $\ln (p_{max}/mc)=14$
for a CR distribution extending from 1GeV to 1PeV.
The condition for magnetic field amplification that
$\int j_{CR}dt=10\sqrt{\rho/\mu_0}$
can be rearranged to give a value for $T_{max}$:
$$
T_{max}= 
0.005 \frac {P_{CR}}{\rho u_s^2}\ 
\rho u_s^3 t
 \sqrt {\frac {\mu_0}{\rho}}
 \eqno{(5)}
 $$
which is equivalent to
$$
T_{max}=8\ n_e^{1/2} u_7^3 t_{1000} \frac {P_{CR}}{\rho u_s^2}
{\rm PeV}
\eqno{(6)}
$$
For characteristic values 
($u_7=0.5$, $n_e=1$, $t=0.4$, $P_{CR}/\rho u_s^2=0.3$)
$T_{max}\sim 100$TeV.
As expected from the above discussion, 
this falls short of the few PeV required to explain Galactic CR.
This will be discussed further in section 8.

Zirakashvili et al (2008a,b) developed a related analytic model of the excitation of the NRH
instability and CR confinement upstream of the shock.
They derived an estimate for the maximum CR energy with
a similar dependency on $n_e^{1/2} u^3 t$ in the numerator of their
equation (19) but with an additional denominator that depends on the magnitude
of the amplified magnetic field.  
Their analysis operates at a more detailed level than ours since they consider
the growth in amplitude of the amplified magnetic field over a range of scales and 
small angle CR scattering by small scale field.
However, the dominant physics is similar in their analysis and ours,
and their estimate of the CR energy (equation (19))
is similar to that in our equation (6).

Note that the magnetic field does not enter into
the estimation of the maximum energy CR in equation (6).
The magnetic field is assumed to grow to whatever
magnitude and spatial scale required to confine CR
at energies less than $T_{max}$.
This assumption is justified on the basis that 
the magnitude of the field increases by a substantial
numerical factor in time $\gamma_{max}^{-1}$ even in the 
non-linear regime and that this is accompanied by
rapid growth in characteristic spatial scale.
For example, the magnetic field and its spatial scale
are very much larger in figure 1 when 
$\int \gamma_{max}dt \approx 8$
than when $\int \gamma_{max}dt \approx 5$
even though the time is different by a factor of only $1.6$.
Allowing the field to grow for just a little longer gives
a much larger magnetic field and spatial scale.
Consequently the primary parameter is not the magnitude of the magnetic field 
but the charge carried by escaping CR.

It is clear that magnetic field amplification is only possible if a population of 
high energy CR escape upstream of the shock.
A magnetic field capable of stopping their escape is only produced after the areal charge $Q_{CR}$
has escaped.
Thus CR escape upstream of the shock is an essential aspect of diffusive shock acceleration
when magnetic field amplification is operational.
If a charge $Q_{CR}$
has not passed through a particular point upstream of the shock
then CR cannot be confined at that distance from the shock.


Previous authors have also concluded that CR escape upstream is inevitable, but generally for different reasons.
One possibility is that CR escape through filamentary cavities in the magnetic field
(Reville \& Bell 2012, 2013)
but usually it is assumed that CR escape at a free escape boundary in position or momentum.
Free escape boundaries were introduced at an early stage in the development of shock acceleration theory
(Ellison et al 1981).
It was appreciated that steady state models were impossible without CR escape
(Ellison \& Eichler, 1984, Berezhko \& Ellison 1999, Malkov et al 2000).
The standard $T^{-2}$ test particle energy spectrum at strong shocks diverges if integrated
to infinite energy. 
The divergence is even stronger if non-linear CR feedback onto the shock structure is included
since the relativistic CR pressure increases the density jump at the shock resulting in 
a spectrum flatter than $T^{-2}$ at high CR energy.
However, the conditions for a steady state solution do not present a compelling argument for a free escape boundary
since time-dependent solutions offer an acceptable option.
Indeed the Lagage \& Cesarsky limit on CR energy is based on the assumption that 
the maximum CR energy increases with time.
The case for CR escape was strengthened by the recognition
that magnetic field amplification is essential 
since there will always be a distance upstream at which amplification is inoperative and CR must escape.
Similarly, there must always be an upper limit to the 
spatial scale to which field is amplified
and a corresponding limit, through the Larmor radius, on the momentum to 
which CR can be accelerated.
This led to the imposition of a free escape boundary 
either at an imposed distance upstream of the shock 
(eg Caprioli et al 2010b, Ohira et al 2010)
or at an imposed CR momentum (eg Ellison \& Bykov 2011).
Whether the free escape boundary is imposed in momentum or configuration space,
the momentum at which CR escape depends on the magnetic field.
Some authors (eg Ptuskin et al 2010) assume that the field is amplified until the magnetic energy density 
reaches a given fraction of $\rho u_s^2$.
The fractional magnetic energy density can be chosen to match observation 
(Ptuskin et al 2010, V\"{o}lk et al 2005).
Alternatively the field can be chosen to match the saturation value estimated by Bell (2004).
Others (Caprioli et al 2010a, Drury 2011) choose a mathematical form for the amplified magnetic field
that allows for multiple possibilites.
Ptsukin \& Zirakashvili (2003) and Caprioli et al (2009a,b, 2010b) include magnetic field generation 
due to the resonant instability as it develops ahead of the shock.
Vladimirov et al (2006) include magnetic field generation
in response to gradients in the CR pressure.
Ellison et al (2012) choose an amplified magnetic field suitable for SNR expansion into
a circumstellar wind.

In contrast to the above, CR escape in our model is determined by the 
escaping CR electric charge
rather than the magnetic field generated in the upstream plasma,
although magnetic field amplification is implicitly required by our model.
Reville et al (2009) also used the escaping flux to calculate growth rates
in an effort to motivate a realistic free-escape boundary location in their steady-state non-linear model. 
Their discussion on self-regulation of CR precursors however 
did not extend to the maximum CR energy.

 
\section{A numerical model}

We now set out to test the above conclusions as far as we are able with a numerical model that 
includes the self-consistent interaction of CR modelled kinetically with 
a background plasma modelled magnetohydrodynamically.
Standard MHD equations describe the background plasma except that a 
$-{\bf j}_{CR}\times {\bf B}$ force is added to the momentum equation:
$$
\rho \frac {d {\bf u}}{d t}=-\nabla P 
-\frac{1}{\mu _0}{\bf B}\times (\nabla \times {\bf B})
-{\bf j}_{CR} \times {\bf B}
\eqno{(7)}
$$
as described in Lucek \& Bell (2000) and Bell (2004).
The CR distribution function $f({\bf r}, {\bf p}, t)$ at position ${\bf r}$ and momentum ${\bf p}$ is defined in the
local fluid rest frame and evolves according to the Vlasov-Fokker-Planck (VFP) equation
$$
\frac {df}{dt}= 
-v_i \frac {\partial f}{\partial r_i}
+ p_i  \frac {\partial u_j}{\partial r_i}\frac{\partial f}{\partial p_j}
- \epsilon _{ijk} e v_i B_j \frac{\partial f}{\partial p_k}
+C(f)
\eqno{(8)}
$$
where $C(f) $ is an optional collision term included to represent scattering by magnetic fluctuations on a small scale.
The electric field is zero in the local fluid rest frame.
Quadratic terms in the local fluid velocity ${\bf u}$ are neglected
on the assumption that $u\ll c$.
The CR current ${\bf j}_{CR}$, required for insertion into the MHD momentum equation,
is calculated by integration over $f$ in momentum space.
The magnetic field in the CR kinetic equation is taken from the MHD calculation.

As in Lucek \& Bell (2000) and Bell (2004) three spatial dimensions are needed
to represent the turbulence adequately.
Since our aim is to investigate the mutual interactions of
magnetic field amplification, CR acceleration and CR escape upstream of the shock
we need to model the complete system including the shock and 
the complete CR precursor.
Because we model the detailed interaction between the CR and the distorted magnetic field
we need to resolve the CR Larmor radius in configuration space
and the rotation of CR trajectories in momentum space.
As a consequence the numerical model should be 6-dimensional in momentum-configuration space
with spatial scales extending from the CR Larmor radius of the lowest energy CR to 
the precursor scaleheight of the highest energy CR.
This would be impossible without extraordinary computational resources
so our strategy is to design a computational model that
includes all the important processes at a minimal level.
We retain the three dimensions in configuration space but limit
the range of CR momentum to a factor of 10 so that we do not have to
resolve the Larmor radius of very low energy CR.
We choose a shock velocity $u_s=c/5$ to keep the ratio of the CR to the MHD timescale to
a minimum while staying close to the range of conceivable SNR expansion speeds.
Our greatest approximations are made in the momentum space representation
of the CR distribution function since this is the aspect of the calculation 
in which the number of dimensions can be reduced.

The Vlasov-Fokker-Planck (VFP) equation (equation 8) is important
in the physics of laser-produced plasmas where it is solved in finite difference form
to model electron transport.
The successful use of VFP simulation to model electron transport
in laser-produced plasmas stretches back more than 30 years (Bell et al, 1981)
so it is natural to apply the techniques to CR which obey the same equation.
The distribution  function $f({\bf r},{\bf p} ,t)$ of charged particles (cosmic rays or energetic electrons) is usually represented in spherical co-ordinates ($p$,$\theta$,$\phi$) in momentum space.
A common representation of the distribution function is as a sum of spherical harmonics:
 \begin{eqnarray*}
 f({\bf r},{\bf p},t)=\sum _{l,m} f_l^m({\bf r},p,t)P_l^{|m|}(\cos \theta)e^{im\phi}
  \hskip 1. cm\\
 \hskip 1.5 cm
 l=0,\infty \ \ \ \ m=-l,l\ \  \ \ f_l^{-m}=\left( f_l^m \right)^*
 \hskip 0.5 cm (9)
 \end{eqnarray*}
 where $f_l^m({\bf r},p,t)$ is the coefficient for the ($l$,$m$) spherical harmonic.
 $f_l^m({\bf r},p,t)$ is function of time, position and magnitude of momentum $p$.
 The spherical harmonics decribe the angular structure on shells
 of constant magnitude of momentum.
 Reviews of the VFP technique, or papers containing a significant review element, are
 Bell et al (2006), Tzoufras et al (2011) and Thomas et al (2012).
 For information on the application of VFP techniques to CR acceleration
 the reader is referred to Reville \& Bell (2013) 
 which includes an appendix setting out the full VFP equations 
 for a spherical harmonic expansion.
 Bell, Schure \& Reville (2012) also apply VFP techniques to CR acceleration.
 The use of an expansion in tensors (used here) as an alternative to
 spherical harmonics
 is discussed  by Schure \& Bell (2012).

 VFP simulation was used by Bell et al (2011) for the calculation
 of CR acceleration by oblique shocks. 
 They found that an expansion to 15th harmonic could be needed for oblique shocks because of the abrupt change in
 magnetic field direction at the shock, but that only a few harmonics are needed for
 quasi-parallel shocks.
Here we reduce the computational size of the problem by modelling parallel shocks in which
the zeroth order magnetic field is parallel to the shock normal.
There are good reasons to suppose that the first few terms in the expansion capture the 
essential physics.
Firstly, as shown by Bell (2004) the NRH instability
is driven by the CR current density ${\bf j}_{CR}$.
Higher order anisotropies do not directly contribute to the instability.
Secondly,
the CR precursor scaleheight is $c/u_s$ times the 
CR scattering mean free path in standard DSA theory.
DSA theory is based on the diffusive approximation in which
only the first order anisotropy is needed
and only the first two terms (isotropy and drift anisotropy)
in the harmonic expansion are retained.
In diffusion theory the higher order anisotropies are damped by scattering.
Hence it might be supposed that only the first two terms are needed and 
an adequate representation of the CR distribution function might be
$f({\bf r},{\bf p},t)=f_0^0({\bf r},p,t)+ f_1^0({\bf r},p,t)\cos \theta+ \Re \{f_1^1({\bf r},p,t) \sin \theta e^{i\phi}\}$.
This `$f_0+f_1$' expansion allows free CR propagation along magnetic field lines but restricts
transport across the magnetic field because the direction of the anisotropic drift term
is rotated by the field.
However, this expansion omits an essential feature of CR transport.
It does not allow CR to gyrate as they travel along a magnetic field line.
The $f_0+f_1$ expansion allows CR to propagate along field lines and separately it allows CR to gyrate around
field lines, but it does not allow CR to do both at the same time.
Spiral trajectories require the inclusion of the off-diagonal components of
the stress tensor $f_2$.
Without the stress tensor, CR cannot resonantly interact in space
with magnetic perturbations on the scale of a Larmor radius.
Clearly this would rule out the Alfven instability,
and more surprising it also rules out the NRH instability.
The role of the stress tensor is discussed by
Schure et al (2011) in which the linear NRH
dispersion relation is derived with the perturbed CR distribution expressed 
as a tensor expansion.
We proceed on the basis that the CR distribution function may be adequately
represented by an isotropic part (0th order in $u_s/c$) plus a 
drift component (1st order in $u_s/c$) plus a term representing 
the stress tensor (2nd order in $u_s/c$).
This second order expansion is more easily represented in 
the equivalent tensor notation
instead of spherical harmonics:
$$
f({\bf r},{\bf p},t)=f_0({\bf r},p,t)
+ f_{i}({\bf r},p,t) \frac{p_i}{p}
+ f_{ij}({\bf r},p,t) \frac{p_i p_j}{p^2}
\eqno{(10)}
$$
where the trace of $f_{ij}$ is zero because it is already accounted for
in the isotropic term $f_0$.
Following Johnston (1960) the reduced Vlasov equation for the CR distribution function is then:
$$
\frac {\partial f_0}{\partial t}+ \frac {\partial (f_0 u_i)}{\partial r_i}
=
-\frac{c}{3} \frac {\partial f_i}{\partial r_i}
+    \frac {\partial u_i}{\partial r_i} \frac {1}{3p^2} \frac {\partial (p^3f_0)}{\partial p}
\eqno{(11a)}
$$
$$
\frac {\partial f_i}{\partial t}+ \frac {\partial (f_i u_j)}{\partial r_j}
=
-c \frac {\partial f_0}{\partial r_i}
- \frac{2c}{5} \frac {\partial f_{ij}}{\partial r_j}
-\epsilon _{ijk} \frac {ceB_j}{p} f_k
\eqno{(11b)}
$$
$$
\frac {\partial f_{ij}}{\partial t}+ \frac {\partial (f_{ij} u_k)}{\partial r_k}
=
-\frac{c}{2} \left ( \frac {\partial f_i}{\partial r_j}
+ \frac {\partial f_j}{\partial r_i} \right )
+\frac{c}{3} \delta _{ij} \frac {\partial f_k}{\partial r_k}
$$
$$
- \frac {ceB_k}{p} \left ( \epsilon _{kil} f_{lj}+ \epsilon _{kjl} f_{li} \right )
\eqno{(11c)}
$$
where quadratic order terms in the velocity ${\bf u}$ have been neglected
and we have omitted terms involving $ \partial f/\partial p$
times a gradient of ${\bf u}$
apart from the term in equation  (11a) for the evolution of $f_0$.
This amounts to the neglect of second order Fermi acceleration
and acceleration resulting from shear motions in the background hydrodynamics.
In our problem, these processes are small in comparison with acceleration at the shock.

We have chosen to terminate the set of equations at equation (11c) by setting $f_{ijk}$
to zero.
Additionally, we soften the termination of the tensor expansion by replacing the 
magnetic rotation in equation (11c) by a damping term with a damping rate equal to the
magnetic gyration frequency.
The logic of this approximation is that random rotation of the stress tensor anisotropy leads to its damping.
A similar assumption underlies Bohm diffusion which replaces rotation of $f_i$ by a
damping rate, thereby terminating the tensor expansion at $f_i$ whereas we terminate it at $f_{ij}$.
Our approximation makes intuitive sense, and we support it in 
Appendices D and E by
examining its effect on propagating modes and on the NRH instability.
\newline
\indent 
The notation is simplified by introducing the vector $g_i$ 
related to shear and vorticity ($\nabla \times {\bf f}_1$) 
in CR motion to represent
the off-diagonal components of the stress tensor.
The on-diagonal components of $f_{ij}$ ($i=j$) are approximately
accounted for by multiplying the 
$-c {\partial f_0}/{\partial r_i}$ term in equation (11b) by 9/5
to allow for the stress tensor contribution to compressional waves
and to allow freely streaming CR to propagate at $\sqrt {3/5}c$ instead of $\sqrt {1/3}c$,
as shown in Appendix D.
The equations to be solved (their derivation is given in Appendix C) are then
$$
\frac {\partial f_0}{\partial t}+ \frac {\partial (f_0 u_i)}{\partial r_i}
=
-\frac{c}{3} \frac {\partial f_i}{\partial r_i}
+    \frac {\partial u_i}{\partial r_i} \frac {1}{3p^2} \frac {\partial (p^3f_0)}{\partial p}
$$
$$
\frac {\partial f_i}{\partial t}+ \frac {\partial (f_i u_j)}{\partial r_j}
=
-\frac{9}{5} c \frac {\partial f_0}{\partial r_i}
-\frac{1}{5}c \epsilon_{ijk}\frac {\partial g_k}{\partial r_j}
-\epsilon _{ijk} \frac{ceB_j}{p} f_k
$$
$$
\frac {\partial g_i}{\partial t}+ \frac {\partial (g_i u_j)}{\partial r_j}
=
c \epsilon_{ijk}\frac {\partial f_k}{\partial r_j}- \nu_B g_i
\eqno{(12)}
$$
Equivalently, expressed in vector notation,
$$
\frac{\partial f_0}{\partial t}+\nabla.({\bf u}f_0)=
-\frac{c}{3}\nabla.{\bf f}_1
+\frac{\nabla.{\bf u}}{3p^2} \frac {\partial (p^3f_0)}{\partial p}
$$
$$
\frac{\partial {\bf f}_1}{\partial t}+\nabla.({\bf uf}_1)=
-\frac{9c}{5}\nabla f_0
-\frac{c}{5} \nabla \times {\bf g}_1
-{\bf \Omega}\times {\bf f_1}
$$
$$
\frac{\partial {\bf g}_1}{\partial t}+\nabla.({\bf ug}_1)=
c \nabla \times {\bf f}_1
-\nu_B {\bf g}_1
\eqno{(13)}
$$
where ${\bf \Omega} = ec{\bf B}/p$ is the vector CR Larmor frequency
and we take $\nu_B=ecB/p$.
The presence of the curls of  ${\bf f}_1$  and ${\bf g}_1$ 
facilitates the propagation of transverse modes in ${\bf f}_1$ and ${\bf g}_1$ 
as needed for helical motion along magnetic field lines
or CR propagation at an angle to the wavevector ${\bf k}$.

The termination of the harmonic expansion at the stress tensor makes the computation tractable with available resources.
Appendices D and E show that the truncated expansion provides an adequate representation
of the essential physics of CR propagation and CR-driven instability.


\section{The Simulation}

A full 3D simulation with realistic parameters is not possible
because of the large ratio of the largest distances 
(the CR precursor scaleheight and the CR free escape distance)
to the smallest distance (the shortest wavelength on which the NRH instability grows).
The correspondingly large ratio of the largest timescale (the SNR expansion time) to
the shortest timescale (the shortest NRH growth time)
similarly makes substantial demands on computer resources.
The computational constraints are discussed in Appendix F.

We artificially increase the magnetic field and stretch other
parameters in a favourable direction, choosing:
$$
n_e=0.1 {\rm cm}^{-3} \hskip 0.2 cm
u_s=60,000 {\rm km \ s}^{-1}\hskip 0.2 cm
$$
$$
B_0=47 \mu{\rm G} \hskip 0.2 cm
v_A=3\times 10^5 {\rm m \ s}^{-1} \hskip 0.2 cm
T_{inject}=100 {\rm TeV} 
\eqno{(14)}
$$
where $T_{inject}$ is the energy at which CR are injected at the shock.
The zeroth order magnetic field is aligned along the shock normal.
The instability is seeded by imposing random fluctuations on the magnetic field
with a mean magnitude of $9\mu{\rm G}$.
CR are injected at the shock into the lowest momentum bin (width $\Delta p$)  according to the rule
$$
4\pi p^2 \Delta p \ cp \frac{\partial f_0 }{\partial t} 
=- {\rm constant} \times \nabla .{\bf u}\ 
\min(\rho_0 u^2 ,U)
\eqno{(15)}
$$
where $\rho_0$ is the density upstream of the shock and 
$u$ is the local background fluid velocity relative to the initially stationary upstream plasma.
$U$ is the local thermal energy density.
This prescription is designed to inject a suitable energy density of CR at the shock, dependent on the choice of the constant, 
whilst avoiding a negative value of $U$ due to excessive transfer of energy to CR from cold plasma at the foot of the shock.
The resulting CR current density ahead of the shock is displayed in panel (e) of figures 2 and 3.
The peak CR current density in the population freely escaping ahead of the shock in figure 2 is
$$
j_{CR} =1.1 \times 10^{-14}{\rm Amp\ m}^{-2}
\eqno{(16)}
$$
This current density corresponds to $\eta= 0.026$, giving
$$
\gamma_{max}^{-1}=2.3 \times 10^6 {\rm sec}\hskip 0.5 cm
c \gamma_{max}^{-1}=7 \times 10^{14} {\rm m}
\eqno{(17)}
$$
$$
k_{max}^{-1}=7 \times 10^{11} {\rm m}\hskip 0.5 cm
r_g=7 \times 10^{13} {\rm m}
$$
$$
r_g k_{max}=100 \hskip 0.5 cm
M_A=200
\eqno{(18)}
$$
We use a spatial grid with $\Delta x= \Delta y = \Delta z = 1.4\times 10^{12} {\rm m}$.
Ten cells in momentum cover an energy range from 100TeV to 1PeV with logarithmic spacing.
There are 32 cells in each of $x$ and $y$ with periodic boundary conditions.
In contrast 5676 cells are used in $z$ with reflective boundary conditions for the MHD part of the code
and for CR at the right hand boundary.
CR reaching the left hand boundary are disposed of on the assumption that they escape freely.
Correspondingly, the computational box extends $7.9\times 10^{15}$m by $4.5\times 10^{13}$m
by $4.5\times 10^{13}$m.
The box-size in $x$ and $y$ is comparable with the initial CR Larmor radius, 
but the Larmor radius contracts significantly as the magnetic field is amplified.
The cell-size is $\pi ^{-1}$ times the wavelength of the fastest growing mode, which is barely sufficient 
to represent the initial growth of the instability, but
the characteristic scalelength of the instability increases rapidly as its amplitude grows.
These parameters are only marginally sufficient to represent the physics,
but for example a halving of the computational cell-size would increase the 
computational cost by a factor of 16.
The simulation in figure 3 was run for $6.2\times 10^7 {\rm sec}=2{\rm yr}$ using 128 processors for 75 hours on the SCARF-LEXICON
cluster at the UK Rutherford Appleton Laboratory.
The MHD part of the code is the same as that used by Lucek \& Bell (2000) and Bell (2004).
The VFP part of the code that models the CR uses a second order Runge-Kutta scheme in configuration space, 
a donor-cell scheme in magnitude of momentum, and the Boris algorithm 
for rotation in momentum due to magnetic field as 
used in particle-in-cell codes (Birdsall \& Langdon 1985). 
The VFP equations are formulated in tensor notation rather than in spherical harmonics,
but the two formulations are similar for the low order expansion used here, 
and their numerical solution
is informed by experience with the KALOS spherical harmonic code (Bell et al, 2006) where Runge-Kutta 
advection is found to be robust, sufficiently accurate, and a good fit to the form of the equations.
Numerical diffusion due to limited spatial resolution is ameliorated
by designing the simulation to initialise the upstream background plasma at rest
relative to the computational grid.
The more usual approach of initiating the simulation by setting the background plasma in motion
towards a reflective boundary would exacerbate the effect of numerical diffusion 
on small structures during advection.
Instead, a dense piston is initialised moving leftward from the right boundary, 
pushing a shock before it into the stationary background plasma.


\section{Simulation results in 3D}

\begin{figure*}
\includegraphics[angle=90,width=12cm]{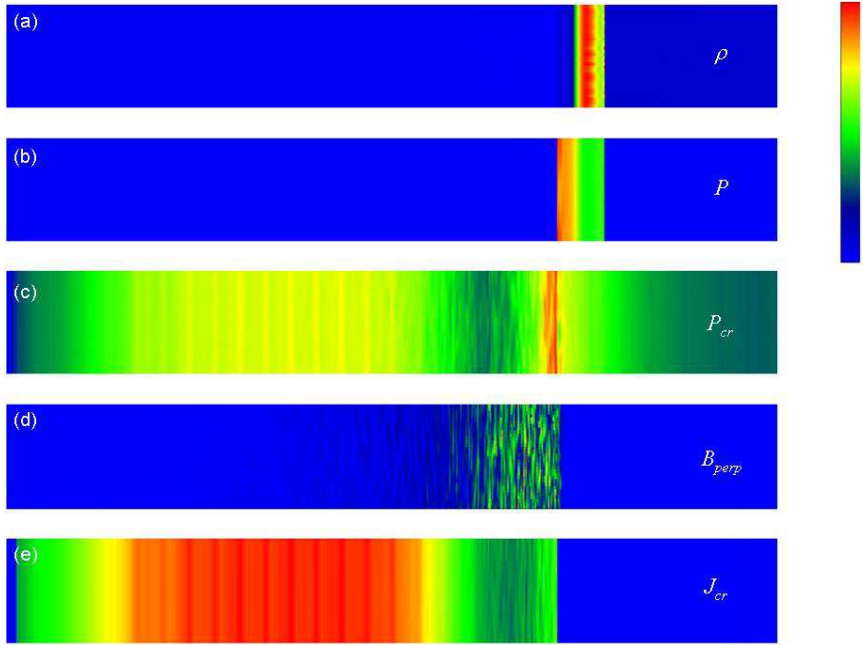}
\caption{
2D slices of a 3D simulation with peak values in brackets: 
electron density ($9.7{\rm cm}^{-3}$), 
pressure (340nPa), 
CR pressure (2.3nPa), 
magnitude of the magnetic field perpendicular to the shock normal (610$\mu$G),
CR current parallel to the shock normal ($1.1\times 10^{-14}{\rm Am}^{-2}$).
$t=4.12\times 10^7{\rm sec}$ (1.3 years).
The dimensions of the computational box are $7.9\times 10^{15}$m by $4.5\times 10^{13}$m.
The horizontal spatial scale, in the direction of the shock normal, is artificially compressed by a 
factor of 24.
}
\label{fig:figure2}
\end{figure*}

\begin{figure*}
\includegraphics[angle=90,width=12cm]{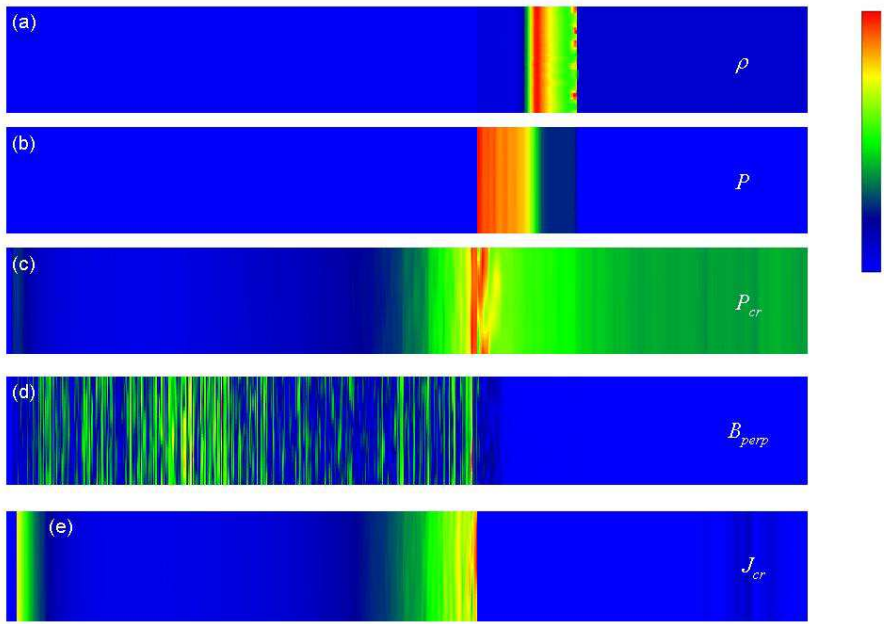}
\caption{
2D slices of a 3D simulation with peak values in brackets:
electron density ($6.9{\rm cm}^{-3}$), 
pressure (320nPa), 
CR pressure (1.9nPa), 
magnitude of the magnetic field perpendicular to the shock normal (590$\mu$G),
CR current parallel to the shock normal ($4.2\times 10^{-15}{\rm Am}^{-2}$).
$t=6.18\times 10^7{\rm sec}$ (2.0 years).
The dimensions of the computational box are $7.9\times 10^{15}$m by $4.5\times 10^{13}$m.
The horizontal spatial scale, in the direction of the shock normal, is artificially compressed by a 
factor of 24.
}
\label{fig:figure3}
\end{figure*}

Results from the 3D simulation are presented in figures 2 and 3
at $t=4.1\times 10^7{\rm sec}$
and $t=6.2\times 10^7{\rm sec}$ respectively.
Note that the horizontal spatial scales, in the direction of the shock normal, 
are artificially compressed by a factor of 24.
The actual aspect ratio of the computational box is 177:1.
The plot of the background plasma density in panel (a) shows the position of
the dense plasma piston propagating leftwards and pushing the shock ahead of it.
The high pressure region in panel (b) is due to plasma heating at the shock.
The CR pressure is plotted in panel (c).
The maximum CR pressure occurs at the shock.
Panel (c) of figure 2 shows the CR population dividing into a 
population freely propagating ahead of the shock and a
population confined by magnetic field (panel (d)) near the shock.
CR confined near the shock continue to be accelerated.
The banded structure in the CR pressure ahead of the shock in panel (c)
is exaggerated by the colour coding.
It represents only a variation of a few percent in the CR pressure
and probably arises from a small oscillation in the injection rate at the shock.
Also, the horizontal spatial compression of figure 2 
distorts the aspect ratio of the bands.

The escaping CR are an essential aspect of the process of CR confinement by the self-generated
magnetic field.
They excite the instability and carry the escaping areal charge 
$Q_{CR}=\int j_{CR}dt$
identified in equation (1) as necessary for substantial field amplification. 
In figure 2 the escaping CR are only just reaching the left hand of the grid
where the integral $\int j_{CR}dt$ is small and field
amplification is negligible.
In contrast, immediately in front of the shock in figure 2  $\int j_{CR}dt$ has reached
the critical value needed for strong field amplification causing the onset of
CR confinement.

By the time of figure 3 most of the escaping population has passed through the
free escape boundary at the left hand end of the grid.
By this time a large magnetic field in the upstream plasma
has switched off CR escape creating an expanse of low CR density between
the confined and escaping CR populations.
Panel (d) shows that the magnetic field is amplified by
an order of magnitude by the escaping CR
over a large distance ahead of the shock.

The separation of CR into escaping and confined populations 
matches expectations from the argument
in section 3.
The CR charge per unit shock area
in the escaping population is $\sim 1\times 10^{-7}$ Coulomb m$^{-2}$ in figure 2
in good agreement with the
estimate of $1.3\times 10^{-7}$ Coulomb m$^{-2}$ from
equation (1).
Panels (e) in figures 2 and 3 plot the CR current
density ahead of the shock.


\section{Simulation results in 2D}

\begin{figure*}
\includegraphics[angle=0,width=14cm]{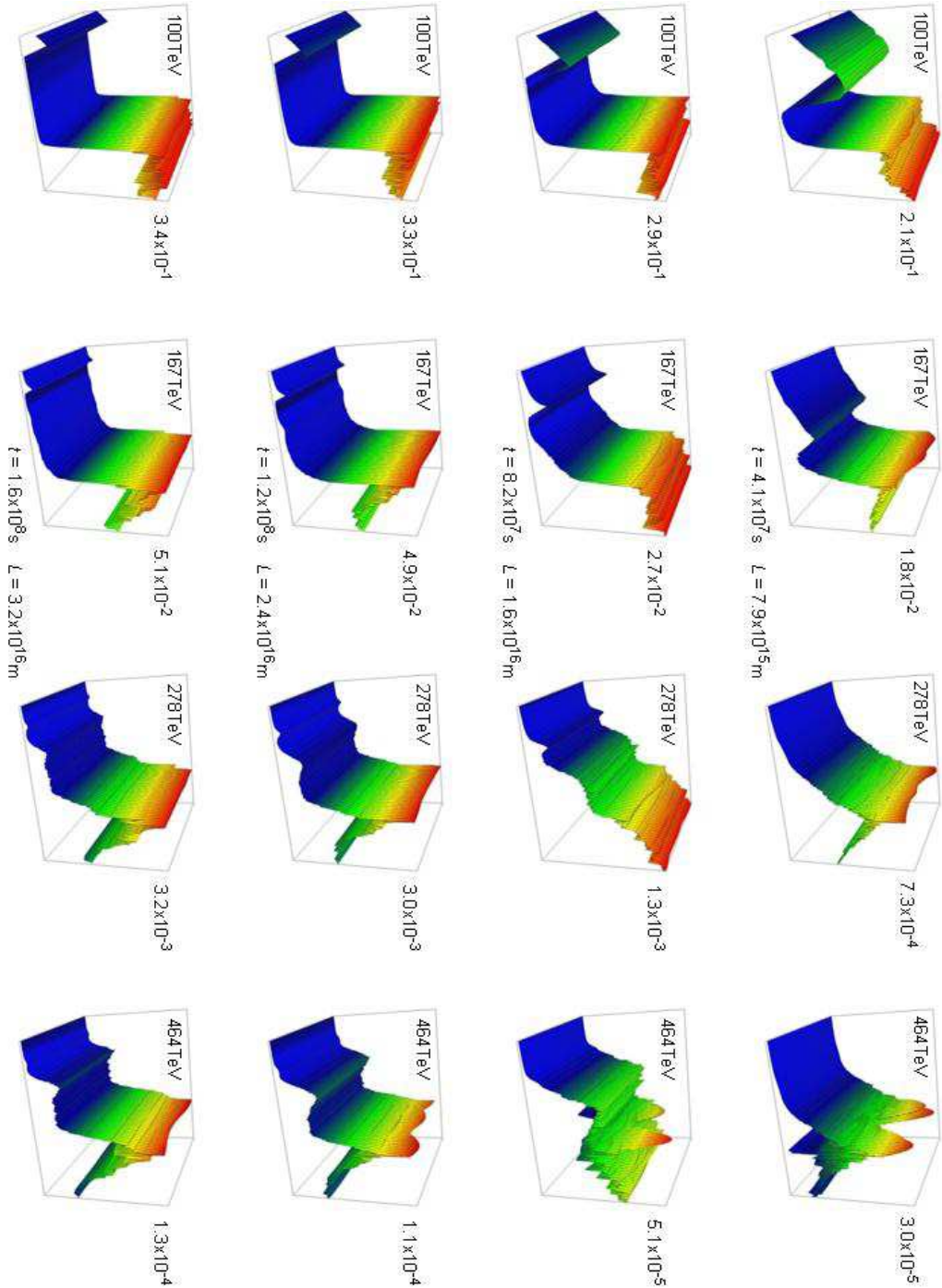}
\caption{
2D simulations. Plots of $f_0p^3$ at CR energies 100, 167, 278 and 464TeV at times between
$4.1\times 10^7{\rm sec}$ (top row)
\& $16 \times 10^7{\rm sec}$ (bottom row),
with computational box lengths $L$ proportional to $t$ between $7.9\times 10^{15}{\rm m}$ 
 \& $3.2 \times 10^{16}{\rm m}$
chosen such that the CR travel the length
of the box during the simulation time.
The width of the computational box is $4.5\times 10^{13}{\rm m}$
in each plot.
The vertical axes are in dimensionless units with peak values given against the axis.
}
\label{fig:figure4}
\end{figure*}

Figure 4 follows the calculations of figures 2 and 3 to later times 
with the same parameters but
in 2D instead of 3D to reduce the demand on computer resources.
The top row of plots in figure 4 is the 2D equivalent of the 3D
results in figure 2 at $t=4.1\times 10^7{\rm sec}$. 
The dimensions of the spatial grid are the same in figure 2 and the top row of figure 4.
In subsequent rows of figure 4 the spatial dimensions normal to the shock are
expanded by factors of 2, 3 and 4 such that CR travel the length of the grid in the respective times
($t=8.2\times 10^7$, $1.2\times 10^8$ \& $1.6\times 10^8{\rm sec}$).

In 3D at $t=4.1\times 10^7{\rm sec}$ the dip in the CR density separating the escaping 
and confined CR has only recently formed as seen in figure 2(c).
At the same time in 2D (figure 4 top row) 
the separation between the 
confined and escaping populations is relatively more developed.
Otherwise the results are similar in 2D and 3D.
The reason for the slightly earlier development of CR confinement in figure 4 is that 
magnetic structures can expand more freely in 2D.
In 2D, magnetic structures expand in the ignored dimension without
running into stationary or counterpropagating plasma.
Nonlinear development in 2D is therefore less inhibited
and the magnetic field grows more rapidly.
The magnetic field in 2D 
reaches a steady value similar to that found in the 3D calculation.
The eventual limit on the magnitude of the magnetic field
may be set by magnetic tension which operates equally in 2D and 3D.

Comparison of the CR distributions at different times in figure 4
shows that at early times ($t=4.1\times 10^7{\rm sec}$) 
only the low energy CR are confined.
At $t=8.2\times 10^7{\rm sec}$ CR are confined at 100 and 167TeV with confinement beginning
to occur at 278TeV and no evidence of the
development of two separate populations in the few CR reaching 464TeV.
By $t=1.2\times 10^8{\rm sec}$ a larger number of CR have reached 464TeV with
a local minimum in CR density $f_0p^3$ evident at all energies up to 464TeV in front of the shock.
By $t=1.2\times 10^8{\rm sec}$ CR at 100 and 167 TeV are strongly confined
with a short precursor scaleheight ahead of the shock, which is unsurprising
since the Larmor radius of a 100TeV proton in a $700\mu{\rm G}$
magnetic field is $5\times 10^{12}{\rm m}$.
By $t=1.6\times 10^8{\rm sec}$ CR are confined at all energies up to 464TeV.

At all four times in figure 4 the CR spectrum is much steeper than the steady-state
test particle spectrum ($f_0 \propto p^{-4}$) because of CR loss into the downstream 
plasma. 
Magnetic field amplification takes place ahead of the shock and 
the downstream field only becomes large
when the shock overtakes the field amplified in the upstream.
Hence at early times CR escape downstream.
This causes a reduction in the number of CR
accelerated to high energy at the shock,
thereby steepening the spectrum.
Downstream confinement at the shock improves at later times as shown
by the downstream gradients in $f_0$ at 
$t=1.2\times 10^8{\rm sec}$ and $t=1.6\times 10^8{\rm sec}$.
However, unphysical numerical relaxation of the 
spatially compressed downstream magnetic field due to 
limited spatial resolution may be playing a part in
allowing CR to escape downstream.

There is slight evidence of pulsed acceleration as seen in the three separate peaks in 
the 278TeV CR density at $t=1.2\times 10^8{\rm sec}$ and $t=1.6\times 10^8{\rm sec}$ 
but overall the CR profiles develop in an orderly manner.


\section{Application to SNR}

Our simulations support the model developed in section 3.
According to our model
CR are confined and accelerated if the  
electrical charge of CR escaping upstream of the shock
reaches $Q_{CR}=10\sqrt{\rho/\mu_0}\ {\rm Coulomb\ m}^{-2}$.
We now apply this to spherical SNR shocks.
The CR current density at a radius $R$
is $ j_{CR}=\eta \rho u_s^3r^2/R^2 T$
due to CR accelerated to energy $eT $ when the shock radius was $r$.
Since only the highest energy CR escape upstream 
we assume that the CR reaching the radius $R$
are monoenergetic
with energy $eT$.
$T$ evolves as the shock expands.
When the SNR shock reaches the radius $R$, CR are confined if
$$
\int _0^R \frac{\eta \rho(r) u_s^2(r)}{T(r)}\ r^2dr
=10R^2\sqrt{\frac{\rho (R)}{\mu _0}}
\eqno{(19)}
$$
Differentiating this equation with respect to $R$ 
and assuming a power law dependence of density on radius,
$\rho(R)=\rho_0 (R/R_0)^{-m}$,
$$T(R)=
\frac{\eta \sqrt{\mu _0}} {5(4-m)} u_s^2 R \sqrt{\rho} 
\eqno{(20)}
$$
Defining $u_7=u_s/10,000{\rm km\ s}^{-1}$, $R_{pc}=R/{\rm parsec}$,
$\eta_{0.03}=\eta/0.03$,
$n_e=\rho/2\times 10^{-21}{\rm kg\ cm}^{-3}$ 
(such that $n_e$ is approximately the electron density in ${\rm cm}^{-3}$),
and taking $m=0$ for expansion into a uniform medium,
$$
T=230\ \eta _{0.03} n_e^{1/2} u_7^2 R_{pc}\ {\rm TeV}
\eqno{(21)}
$$
A SNR with $u_7=0.6$, $n_e=1$ and $R_{pc}=1.7$,
representative of Cas A,
would then accelerate CR to $\sim $140TeV.
A SNR with $u_7=0.5$, $n_e=0.1$ and $R_{pc}=10$,
representative of SN1006,
would accelerate CR to $\sim $180TeV.
These energies are a factor of 10 lower than the energy of the knee.
Abbasi et al (2012) place the knee at 4-5PeV, although data from other experiments
indicate a lower energy and the turnover in the spectrum 
is not well defined as shown in their figure 15.

Our model places a considerable question mark over the ability
of the well-known historical SNR to accelerate CR to the knee.
Acceleration by SNR such as Cas A and SN1006 fails to reach
the knee in our analysis because their expansion is already significantly decelerated.
Zirakashvili \& Ptuskin (2008a) also reach the conclusion (their Table 2) that the historical SNR
do not accelerate CR beyond 100-200 TeV, but note that their parameter 
$\eta _{esc}$ differs from our parameter $\eta$ by a factor of two
($\eta _{esc}=2\eta$).

Initial expansion velocities of very young SNR can reach 
$30,000\ {\rm km\ s}^{-1}$ (Manchester et al 2002)
or possibly higher for some types of SN (Chevalier \& Fransson 2006).
According to equation (21)
expansion into a density of $1\ {\rm cm}^{-3}$
at $30,000\ {\rm km\ s}^{-1}$ for 16 years to
a shock radius of 0.5 parsec  
accelerates CR to $\sim 1{\rm PeV}$.
Moreover, the energy processed through the shock is
comparable to that for expansion to 1.5 parsec
at $6,000\ {\rm km\ s}^{-1}$, thereby contributing a comparable CR energy content
to the Galactic energy budget.
A complementary perspective on the same problem is obtained
by expressing the maximum CR energy in terms of
the mass $M=4\pi \rho R^3/3$ swept up by the shock
and the characteristic energy of the blast wave
$E=Mu_s^2/2$.
If $M_\odot$ is the mass in units of a solar mass
and $E_{44}$ is the energy in units of $10^{44}$J,
the maximum CR energy is
$$
T= 0.5 \eta _{0.03} n_e^{1/6} E_{44} M_\odot ^{-2/3}\ {\rm PeV}
\eqno{(22)}
$$
which indicates that a typical SNR should be able to accelerate CR to $\sim 1 {\rm PeV}$
and that the maximum CR energy is greater for the same energy $E$ given to a smaller mass $M$.
The maximum CR energy is nearly independent of density $n_e$
because a shock expands to a larger radius in a low density medium
before deceleration (Hillas 2006).

In the self-similar Sedov phase,
$E$ is constant, $M$ is proportion to $R^3$ and hence
$T \propto R^{-2}$.  The maximum CR energy 
decreases with radius during the Sedov phase until magnetic field amplification
ceases, at which point our analysis in terms of the charge carried by 
escaping CR is inapplicable.

Equation (20) with $m=2$ indicates that the highest CR energies might be
achieved by SNR shocks expanding into a dense circumstellar medium
previously ejected as a wind from the SN progenitor.
$\rho R^2$ is independent of radius in a steady wind, 
in which case the CR energy
depends only on $u_s^2$  and $\eta$ according to equation (20)
thus favouring CR acceleration in the very early stages of rapid expansion
as previously suggested by V\"{o}lk \& Biermann (1988)
and Bell \& Lucek (2001).
The maximum CR energy for a SNR expanding 
into a wind carrying  
$\dot{M}_5 \times 10^{-5}$ solar masses per year 
and shed with a velocity of $v_{4}\times 10\ {\rm km\ s}^{-1}$ is
$$
T=760 \eta_{0.03} u_7^2 \sqrt{\frac{\dot{M}_5}{v_4}}\ {\rm TeV}
\eqno{(23)}
$$
indicating that PeV energies are attainable.
CR may be accelerated to energies beyond the knee
if the initial shock velocity is $\sim 30,000\ {\rm km\ s}^{-1}$
and the shock expands into a particularly dense wind
or otherwise dense circumstellar medium.

The estimates made in this section are subject to considerable numerical uncertainty.
For example our estimates for the efficiency $\eta \sim 0.03$ 
or the escaping charge 
$Q_{CR}=10\sqrt{\rho/\mu_0}$
could be uncertain by a factor of 2 or 3.
However, the accumulated error in our estimates would need to be a factor of 10 
for the well-known historical SNR
to account for acceleration to the knee.
Our arguments are not completely watertight but
we tentatively conclude that acceleration to the knee takes place in younger
relatively undecelerated SNR.


\section{The CR energy spectrum}

In this section we discuss the energy spectrum of escaping CR
integrated over the lifetime of the SNR.
The energy $T$ of escaping CR changes as the SNR evolves in radius and expansion velocity.
The integrated energy spectrum of CR escaping into the interstellar medium (ISM)
need not be the same as the spectra
of CR at the shock during acceleration or that of CR carried downstream into the interior of the SNR.
Related analyses based on different models of CR escape can be found in 
Caprioli et al (2010a), Drury (2011), Ohira et al (2010),
Ptuskin \& Zirakashvili (2003) and Ptuskin et al (2010).

We assume a power-law density gradient, $\rho \propto R^{-m}$ 
where $m=0$ for a uniform circumstellar medium
and $m=2$ for a steady pre-supernova wind.
We further assume that the shock velocity can be approximated as a power law
$u_s \propto R^{-q}$
over a sufficiently large part of the SNR's evolution.
In this approximation $u_s \propto t^{-q/(1+q)}$
and $R\propto t^{1/(1+q)}$.  In the Sedov phase $q=3/2$.
From equation (20)
$T\propto R^{1-2q-m/2}$.

In a uniform circumstellar medium ($m=0$) the energy $T$ of escaping CR
decreases during expansion if $q>1/2$, 
which is equivalent to $u_s$ decreasing more rapidly than $t^{-1/3}$.
If a SNR expands into a steady pre-supernova wind ($m=2$) the maximum CR energy 
always decreases with time provided $u_s$ decreases with time ($q>0$) as expected.

Let $\int _T^{\infty} E(T)dT$ be the total energy given to CR above an energy $T$.
By definition of $\eta$ the CR energy flux escaping upstream is $\eta \rho
u_s^3$, so 
$\int _T^{\infty} E(T)dT=  \int_0^R \eta 4\pi r^2 \rho u_s^2 dr$
and
$$
E=-\frac{dR}{dT}\  \eta 4\pi R^2 \rho u_s^2
\eqno{(24)}
$$
where CR with energy $T$ escape when the the shock radius is $R$.
From equation (20)
$$
\frac{T}{T_0}=\left ( \frac{R}{R_0} \right )^{1-2q-m/2}
\hskip 0.25 cm
{\rm where}
\hskip 0.25 cm
T_0=\frac{\eta \sqrt{\mu _0}}{5(4-m)} u_0^2 R_0 \sqrt{\rho_0}
\eqno{(25)}
$$
and $\rho_0$, $u_0$ and $T_0$ are the values of 
$\rho$, $u_s$ and $T$ at a reference radius $R=R_0$.
Manipulating these equations gives
$$
N_{CR}=\frac{4\pi \eta \rho_0 u_0^2 R_0^3}{(2q-1+m/2)eT_0^2}
\left ( \frac{T}{T_0} \right )^{-\alpha}
$$
$$
{\rm where}
\hskip 0.2 cm
\alpha=\frac{4q+2}{4q+m-2}
\hskip 1 cm
\eqno{(26)}
$$
and $N_{CR}(T)=E/eT$ is the CR differential spectrum in energy.
For SNR expansion into a uniform medium the CR spectral index is
$
\alpha _{uniform}=(2q+1)/(2q-1)
$,
and for expansion into a wind the index is
$
\alpha _{wind}=(2q+1)/{2q}
$.
During the Sedov phase $m=0$ and $q=3/2$, giving
$$
\alpha _{Sedov}=2
\eqno{(27)}
$$
although the analysis only applies 
while the magnetic field is being amplified by the NRH instability.
A slightly less rapid decrease in shock velocity,
$u_s \propto t^{-0.57}$ ($q=4/3$), would reproduce
the CR spectral index ($\alpha \approx 2.2$)
inferred for CR at their source at energies less than 1PeV
(Gaisser et al 1998, Hillas 2005).
The spectral index of CR escaping in the Sedov phase, $\alpha_{Sedov}=2$,
is the same as that for test particle acceleration at a strong shock.
There is no obvious reason why this should be so.

We emphasise that this discussion and the derived spectral index $\alpha$
or $\alpha _{Sedov}$ applies only to CR escaping upstream from the shock 
during SNR expansion.
A further population of lower energy CR are carried into the centre of the SNR
where they reside until they are released into the interstellar medium when the
SNR slows, disintegrates and dissolves into its surroundings.
These lower energy particles lose energy adiabatically as the SNR expands
but they can be expected to contribute most of the Galactic population
of low energy CR.

In section 8 we suggested that acceleration beyond the knee
might result from high velocity expansion into a dense circumstellar medium.
The CR population beyond 1PeV is an uncertain mix of protons and heavy ions.
The overall spectral index of CR released into the Galaxy by SNR beyond 1PeV
can be approximated as
$\alpha \approx 2.7$
with further spectral steepening occurring during propagation
from the source to the Earth.
This spectral index at source is predicted by our analysis if
$q=0.29$ ($u_s \propto t^{-0.22}$) for expansion into a steady wind.
It would correspond to a reduction in the shock velocity 
by a quite reasonable factor of 2.1 between
$t=10\ {\rm years}$  and $t=300\ {\rm years}$.
Expansion into a uniform medium would require
$q=1.09$ ($u_s \propto t^{-0.52}$) equivalent to a reduction in shock velocity
by a less reasonable, but not impossible, factor of 5.9 between
$t=10\ {\rm years}$  and $t=300\ {\rm years}$.
This supports the contention that acceleration to $10-100{\rm PeV}$ may occur at a
SNR shock expanding into a circumstellar wind.


\section{Magnetic field}

The Hillas parameter $\xi u_sBR$ (shock velocity times magnetic field times spatial scale)
provides an estimate of the energy $T_{max}$ to which CR can be accelerated under various circumstances
(Hillas 1984).  
$\xi$ is a numerical factor of order unity, probably lying in the range between 1/8 and 3/8 
depending upon the CR diffusion coefficient
(Lagage \& Cesarsky 1983a,b, Bell 2012).
$T_{max}=4\xi_8 u_7B_{\mu G}R_{pc} {\rm TeV}$ 
where $\xi = \xi_8 /8$ ($\xi_8 =1$ for $\xi=1/8$).
Combining this with equation (21) gives an estimate of the pre-shock magnetic field
required to accelerate CR to $T_{max}$:
$$
B\sim 60 \eta_{0.03} \xi_8^{-1} n_e^{1/2} u_7 \ \mu {\rm G}\ .
\eqno{(28)}
$$
The post-shock field is $\approx 3 \times $ larger due to compression at the shock. 
For our simulation parameters ($n_e=0.1$, $u_7=6$) the estimated post-shock maximum magnetic field is 
$\sim 400 \mu {\rm G}$ which is consistent with the field of $\sim 600 \mu{\rm G}$ seen
in panel (d) of figures 2 and 3.
This confirms that the magnetic field in the simulation is amplified sufficiently
to confine and accelerate CR and that diffusion in the CR precursor
is approximately Bohm, $D\sim r_gc$.

The above estimate of the magnetic field is derived on the basis that the magnetic field must be sufficient
to contain and accelerate CR to the energy estimated in equation (21).
Continued growth in the magnetic field would inhibit CR escape 
and remove the CR current that drives the NRH instability.
Part of the energy generated by the NRH instability is stored in the
kinetic energy of plasma motions.
These motions might continue to stretch magnetic field lines and further increase the magnetic field
after the CR current becomes inhibited.
Further release of CR into the upstream plasma would then be heavily restricted until
the magnetic field relaxes to a lower level.
This might result in oscillation about a marginal state
defined by a balance between magnetic field amplification and CR
escape upstream.
Weak evidence for periodic releases of CR into the 
upstream plasma can be found in the plot of
the 278TeV CR density at $1.2\times 10^8{\rm sec}$ 
in figure 4 but on the whole the system
appears to evolve without oscillation.

In planar geometry escaping CR are in principle capable of 
generating magnetic field at an unlimited distance ahead of the shock.
In the spherical geometry of an expanding SNR 
the CR current decreases with distance ahead of the shock, ${\bf j}_{CR}\propto R^{-2}$, 
so continuous CR escape is needed to amplify the magnetic field at 
a general radius $R$
before the shock reaches that point.
Hence the marginal balance between CR escape and magnetic field generation is more likely in spherical
than planar geometry.

The above discussion assumes that magnetic field growth and CR acceleration
is determined by the growth rate of the NRH instability.
However, it is possible that the instability might saturate and stop growing before
it reaches that given by equation (28).
Bell (2004, 2009) argues that 
tension in the field lines limits amplification when
$\nabla \times {\bf B}\sim \mu_0 {\bf j}_{CR}$
for magnetic field structured on the scale of a CR Larmor radius.
This implies a saturation magnetic energy density
$B_{sat}^2/\mu _0\sim j_{CR}T/c$ and predicts a saturated upstream magnetic field 
$$
B_{sat} \sim 160 \ \eta _{0.03}^{1/2} n_e^{1/2} u_7^{3/2} \ \mu {\rm G}
\eqno{(29)}
$$
with a further $\sim 3\times$ increase at the shock.
The ratio of the magnetic field given by equation (28) to the saturated magnetic
field is
$$
\frac {B} {B_{sat}}\sim  0.4 \xi_8^{-1} \eta_{0.03}^{1/2} u_7^{-1/2}
\eqno{(30)}
$$
which implies that tension in the magnetic field does not
stop the field growing to that given in equation (28).
However, it would not require magnetic field growth to overshoot that
indicated by equation (28) by a large factor before the magnetic tension intervenes
to halt growth.
According to equation (28) the magnetic energy density is proportional to $\rho u_s^2$,
whereas the magnetic energy density determined by saturation is proportional to $\rho u_s^3$.
Observations of SNR slightly favour a dependence on $\rho u_s^3$, 
but the difference is too close to be called (Vink 2006).

From equation (20) ($T=0.05 \eta u_s^2 R  (\rho \mu_0)^{1/2}$ for $m=0$) and the modified Hillas condition ($T=u_sBR/8$)
the required post-shock magnetic energy density is
$$
\frac{B^2}{2\mu_0}\sim \eta ^2 \rho u_s^2\ .
\eqno{(31)}
$$
For magnetic field structured on the scale of the Larmor radius of the highest energy CR
we should assume $\eta \sim 0.03$ as the fraction of $\rho u_s^2$ given to the highest energy CR
in which case the the post-shock magnetic energy density would be 
$\sim 0.1 \%$ of $\rho u_s^2$,
allowing for compression at the shock.
V\"{o}lk et al (2005) find observationally that the post-shock magnetic energy density is typically 
$\sim 3 \%$ of $\rho u_s^2$ in the historical SNR. 
However, magnetic field will also be amplified on the scale of the Larmor radii of low energy 
as well as high energy CR.
The difference between $\sim 0.1 \%$ and $\sim 3 \%$
may be explained by integration over the magnetic structures on scales varying
by six orders of magnitude corresponding to the difference between the Larmor radii of GeV and PeV protons.
If this is the case, most of the magnetic energy at the shock resides at scalelengths 
too short to accelerate CR to PeV energies.
The magnetic field inferred from x-ray synchrotron observations of a SNR shock should not be 
inserted without adjustment into the Hillas parameter to estimate the maximum CR energy.

Our analysis of CR escape leads to the result that the energy density of the magnetic field
confining the highest energy CR is proportional to $\rho u_s^2$.
Previous analyses of CR escape often start from the assumption that the magnetic energy density
is proportional to  $\rho u_s^2$ and consequently they produce similar results for
the spectrum of escaping CR.
Comparable results for the CR spectrum produced in the Sedov phase can be found in
Caprioli et al (2010a), Drury (2011), Ohira et al (2010),
Ptuskin \& Zirakashvili (2003) and Ptuskin et al (2010).
For example a $T^{-2}$ energy spectrum for escaping CR in the Sedov phase has previously been derived by
Berezhko \& Krymskii (1988), Ptuskin \& Zirakashvilii (2005),
Caprioli et al (2010a) and Drury (2011).


\section{CR energy input to the Galaxy}

In our model only the highest energy CR escape upstream of the shock.
At any given time more CR energy is carried away downstream into the SNR than 
escapes into the Galaxy.
Efficient production of Galactic CR might therefore seem impossible.
However, the CR energy carried into the interior of the remnant is subsequently recycled to
drive the SNR expansion and is available to accelerate further CR at a later stage.
The overall efficiency of the production of Galactic CR is demonstrated by integrating 
over the lifetime of the SNR.
Assuming a $T^{-2}$ energy spectrum ($\alpha =2$, $q=3/2$) equation (26)
can be integrated to deduce a total CR energy input to the Galaxy:
$$
E_{total}= \frac{ 3 \eta }{2} \frac {4\pi R_0^3}{3} \rho_0 u_0^2 \ln \left ( \frac {T_2}{T_1} \right)
\eqno{(32)}
$$
between energies $T_1$ and $T_2$ ($\sim 1 {\rm GeV}$ and $\sim 1 {\rm PeV}$ respectively).
A small value of $\eta$ ($\sim 0.03$) is balanced by the factor $\ln (T_{2}/T_{1})$
which results from recycling CR energy carried into the interior of the SNR.

It is occasionally remarked that CR acceleration by very young SNR during the
first few years cannot inject sufficient energy into the
Galaxy to account for CR at energies beyond the knee because the SNR shock is small.
However, if the CR spectrum connects smoothly across the knee and the spectrum 
beyond the knee of escaping CR 
matches observation as shown to be possible in section 9
then it follows that the CR energy
input is sufficient to match the Galactic energy budget.

\section{Conclusions}

The central message of this paper is that CR of a given energy escape freely ahead of a shock
until magnetic field amplification takes place to inhibit propagation.
The condition for propagation inhibition is that a sufficient number of CR must escape upstream
for the NRH instability to grow through $\sim 5-10$
e-foldings at the growth rate of the fastest growing mode.
Since the instability is driven by the CR current, the condition is that a CR electric charge
$Q_{CR} \sim 10\sqrt{\rho /\mu_0}$ per unit area must escape through a spherical surface surrounding the SNR
to amplify the magnetic field and inhibit CR escape through that surface.
Since high energy CR carry less charge than  low energy CR for a fixed CR energy flux 
the condition on $Q_{CR}$ determines the energy of escaping CR.
We find that the energy $eT$ of escaping CR is proportional to $\eta R u_s^2 \sqrt{\rho}$
as given by equation (21).
The energy $eT$ varies during the evolution of the SNR and determines
 the energy spectrum of CR injected into the interstellar medium by SNR.
In our estimation, the historical SNR (Cas A, Tycho, Kepler, SN1006) are currently 
accelerating CR to
$\sim 100-200$ TeV. 
Acceleration to the knee at a few PeV takes place in SNR at an earlier stage of evolution when
the shock velocity is $\sim 10,000\ {\rm km\ s}^{-1}$ or greater.
This is an unsurprising conclusion since if the historical SNR were to accelerate CR to the knee
we would be asking why even higher energy CR were not being produced by 
younger SNR.
Observations by the planned Cherenkov Telescope Array (CTA)
should be crucial in testing our conclusions (Hinton \& Hofman 2010, Aharonian 2012).

Acceleration beyond the knee may take place in very young SNR expanding at $20-30,000\ {\rm km\ s}^{-1}$
into a dense circumstellar pre-supernova wind.

The spectral index of escaping CR is consistent with the measured Galactic CR spectrum at energies less than 1PeV.
Beyond the knee the proton spectral index is uncertain both theoretically and observationally.
The theoretical prediction depends on the rate at which the SNR shock decelerates
during its early expansion.

The magnetic field can be estimated from the Hillas parameter 
as the field needed to accelerate CR to the escape energy.
The field is close to, but slightly less than,
the saturation field determined by tension in the magnetic field.
The predicted magnetic fields are consistent with those observed in SNR
if allowance is made for the large range of scalelengths,
corresponding to the range of CR Larmor radii,
in the structure of the magnetic field.
        
The model is tested against numerical simulation.
The MHD/VFP code is three-dimensional in space and 
one dimensional in CR momentum with anisotropy included to second order.
The computational parameters are pushed to their limit
to allow solution of this multi-scale multi-dimensional problem 
but the results support the analytic model.
In the simulation CR are seen to escape upstream with the electrical charge
predicted by theory
and magnetic field is strongly amplified to the predicted level.

\section{Acknowledgements}

The research leading to these results has received funding
from the European Research Council under the European
Community's Seventh Framework Programme (FP7/2007-
2013) / ERC grant agreement no. 247039 and from grant number ST/H001948/1
made by the UK Science Technology and Facilities Council.
We thank the STFC's e-Science facility for access to the SCARF computing cluster.
We thank an anonymous referee for comments that considerably improved the presentation of this paper.


\section{References}

Abbasi R. et al, 2012, submitted to Astropart Phys, arXiv.1207.3455
\newline
Aharonian F.A., 2012, Astropart Phys, available on-line: 10.1016/j.astropartphys.2012.08.007
\newline
Axford W.I., Leer E. \& Skadron G., 1977, Proc 15th Int. Cosmic Ray Conf., 11, 132
\newline
Bell A.R., 1978, MNRAS, 182, 147
\newline
Bell A.R., 2004, MNRAS, 353, 550
\newline
Bell A.R., 2005, MNRAS, 358, 181
\newline
Bell A.R., 2009, Plasma Phys Cont Fusion, 51,124004
\newline
Bell A.R., 2012, Astropart Phys, available on-line: 10.1016/j.astropartphys.2012.05.022
\newline
Bell A.R., Evans R.G. \& Nicholas D.J., 1981, Phys Rev Lett 46, 243 
\newline
Bell A.R. \& Lucek S.G., 2001 MNRAS 321 433
\newline
Bell A.R., Robinson A.P.L., Sherlock M. Kingham R.J. \& Rozmus W.,
2006, Plasma Phys Cont Fusion, 48, R37
\newline
Bell A.R., Schure K.M., Reville B., 2011, MNRAS 418, 1208
\newline
Berezhko E.G. \& Ellison D.C., 1999, ApJL 526, 385
\newline
Berezhko E.G. \& Krymskii G.F., 1988, Sov Phys Usp 31, 27
\newline
Berezhko E.G., Ksenofontov L.T. \& V\"{o}lk H.J., 2003, A\&A 412, L11
\newline
Birdsall C.K. \& Langdon A.B., 1985, {\it Plasma Physics via Computer Simulation},
McGraw Hill, Singapore.
\newline
Blandford R.D. \&  Ostriker J.P., 1978, ApJ, 221, L29
\newline
Bykov A.M., Osipov S.M. \& Ellison D.C., 2011, MNRAS 410, 39
\newline
Caprioli D., Blasi P. \& Amato E., 2009a, MNRAS 396, 2065
\newline
Caprioli D., Blasi P., Amato E. \& Vietri M., 2009b, MNRAS 395, 895
\newline
Caprioli D., Blasi P. \& Amato E., 2010a, Astropart Phys 33, 160
\newline
Caprioli D., Blasi P. \& Amato E., 2010b, Astropart Phys 33, 307
\newline
Chevalier R.A. \&  Fransson C., 2006,
ApJ, 651, 381
\newline
Drury L.O'C., 2011, MNRAS 415, 1807
\newline
Drury L.O'C. \& Downes T.P., 2012, MNRAS in press, arXiv:1205.6823 
\newline
Drury L.O'C. \& Falle S.A.E.G., 1986, MNRAS 222, 353.
\newline
Ellison D.C. \& Bykov A.M., 2011, ApJ 731, 87
\newline
Ellison D.C. \& Eichler D., 1984, ApJ 286, 691
\newline
Ellison D.C., Jones F.C. \& Eichler D., 1981, J Geophys 50, 110
\newline
Ellison D.C., Slane P., Patnaude D.J. \& Bykov A.M., 2012, ApJ 744, 39
\newline
Gaisser T.K. Protheroe R.J. \& Stanev T., 1998, ApJ 492, 219
\newline
Hillas A.M., 1984, ARA\&A 22, 425
\newline
Hillas A.M., 2005, J Phys G 31, R95
\newline
Hillas A.M., 2006, J Phys Conf Ser 47, 168
\newline
Hinton J.A. \& Hofmann W., 2010, Ann Rev Astron Astrophys, 47, 523
\newline
Johnston T.W., 1960, Phys Rev 120, 1103
\newline
Krymskii G.F., 1977, Sov Phys Dokl, 23, 327
\newline
Kulsrud R. \& Pearce W.P., 1969, ApJ 156 445
\newline
Lagage O. \& Cesarsky C.J., 1983a, A\&A 118 223
\newline
Lagage O. \& Cesarsky C.J., 1983b, ApJ 125 249
\newline
Lucek S.G. \& Bell A.R., 2000, MNRAS 314 65
\newline
Malkov M.A. \& Diamond P.H., 2009, ApJ 692, 1571
\newline
Malkov M.A., Diamond P.H. \& V\"{o}lk H.J., 2000, ApJ 533, L171
\newline
Manchester R.N., Gaensler B.M., Wheaton V.C., Staveley-Smith L., 
Tzioumis A.K., Bizunok N.S., Kesteven M.J.  \& Reynolds J.E.,
2002,
Pub.~Astr.~Soc.~Austr., 19, 207
\newline
Ohira Y., Mirase K. \& Yamazaki R., 2010, A\&A 513 A17
\newline
Ptuskin V.S. \& Zirakashvili V.N., 2003, A\&A 403, 1
\newline
Ptuskin V.S. \& Zirakashvili V.N., 2005, A\&A 429, 755
\newline
Ptuskin V.S., Zirakashvili V.N. \& Seo E.-S., 2010, ApJ 718, 32
\newline
Reville B. \& Bell A.R., 2012, MNRAS 419, 2433
\newline
Reville B. \& Bell A.R., 2013, MNRAS in press, arXiv:1301.3173
\newline
Reville B., Kirk J.G. \& Duffy P., 2009, ApJ 694, 951
\newline
Rogachevskii I., Kleeorin N., Brandenburg A. \& Eichler D., 2012, ApJ 753:6
\newline
Schure K.M. \& Bell A.R., 2011, MNRAS 418, 782
\newline
Schure K.M., Bell A.R., Drury L.O'C. \& Bykov A.M., 2012, Space Science Reviews 173, 491
\newline
Stage M.D., Allen G.E., Houck J.C. \& Davis J.E., 2006, Nature Physics 2, 614
\newline
Thomas A.G.R., Tzoufras M., Robinson A.P.L., Kingham R.J., Ridgers C.P., Sherlock M., \& Bell A.R.,
2012, J Comp Phys 231, 1051 
\newline 
Tzoufras M., Bell A.R., Norreys P.A. \& Tsung F.S.,
2011, J Comp Phys 230, 6475 
\newline
Uchiyama Y. et al, 2007, Nature 449, 576
\newline
Vink J. 2006, Proc of the The X-ray Universe 2005 (ESA SP-604) p319. ed. A. Wilson
\newline
Vink J. \& Laming J.M., 2003, ApJ, 584, 758
\newline
Vladimirov A., Ellison D.C. \& Bykov A.,  2006, ApJ 652, 1246
\newline
V\"{o}lk H.J., Berezhko E.G. \& Ksenofontov L.T., 2005, A\&A, 433, 229
\newline
V\"{o}lk H.J. \& Biermann P., 1988 ApJL 333, L65
\newline
Weibel E.S., 1959, Phys Rev Lett 2, 83
\newline
Wentzel D.G., 1974, ARA\&A 12, 71
\newline
Zirakashvili V.N. \& Ptuskin V.S., 2008a,  ApJ 678, 939
\newline
Zirakashvili V.N., Ptuskin V.S. \& V\"{o}lk H.J., 2008b,  ApJ 678, 255
\newline


\vskip 0.3 cm
\noindent
{\bf Appendix A: The value of $\eta$}

The CR electric current ${\bf j}_{CR}$ drives the amplification of magnetic field through the NRH instability.
Throughout this paper we express  $j_{CR}$ as a fraction of the CR current needed to
carry the characteristic energy flux $\rho u_s^3$:
$j_{CR}=\eta \rho u_s^3/T$ where $T$ is the characteristic CR energy in eV.
In this Appendix we briefly explain why we choose $\eta \approx 0.03$
as our best estimate (see also Bell 2004).

In the absence of CR acceleration the thermal energy density downstream of a strong shock
is $9\rho u_s^2/8$ from the Rankine-Hugoniot relations.
Assuming that a third of this energy is given to CR 
as required for efficient CR production by SNR,
the downstream CR energy density 
is $3\rho u_s^2/8$.
From continuity across the shock the CR energy density immediately ahead of the shock
is also $3\rho u_s^2/8$
and the CR energy flux relative to the upstream plasma is $3\rho u_s^3/8$.
However, only the highest energy CR escape upstream.
Lower energy CR do not penetrate far upstream and they amplify magnetic field on too small 
a scale to engage the escaping CR.
Hence the analysis in this paper depends on the current carried 
only by high energy CR.
For a $T^{-2}$ CR energy spectrum extending from $T_{min}\approx 1{\rm GeV}$ to $T_{max}\approx 1{\rm PeV}$
the energy is spread equally across each decade in energy
with a fraction $1/\ln (T_{max}/T_{min})\approx 1/14$ associated with any energy $T$.
The energy flux carried by CR with energy $T$ is then 
$\sim 3\rho u_s^3/8\ln (T_{max}/T_{min}) \sim 0.03 \rho u_s^3$.
The energy flux carried by CR streaming at velocity $v$ with number density $n_{CR}$ and energy $eT$
is $n_{CR}veT=j_{CR}T$.  
Consequently we assume that 
$j_{CR}=\eta \rho u_s^3/T$ where $\eta \sim 0.03$.


\vskip 0.3 cm
\noindent
{\bf Appendix B: The derivation of equation 2}

Equation 2 for the electric current $j_{CR}$ carried by escaping CR assumes 
(i) that CR escape upstream in a small range of momenta just above $p_{max}$
(ii) that any CR reaching a momentum $p_{max}$
or higher freely stream ahead of the shock at a velocity much greater than the shock velocity $u_s$ 
(iii) that escape is predominantly upstream.
The electric current density is
$$
j_{CR}=\int _{pmax}^\infty \frac{4\pi}{3}f_1 cp^2 dp
\eqno{(B1)}
$$
where $f_1 p_r/p$ is the anisotropic part of the distribution fucntion representing CR drift in the $r$ direction.
In a steady state on an acceleration timescale $\partial f_0/\partial t$ can be neglected from equation 11a.
The second term $\partial (f_0 u_i)/ \partial r_i$ can also be neglected since the CR are assumed to be freely streaming 
at energies above $cp_{max} $ and
advection at the fluid velocity is small.  
In one spatial dimension $r$ equation 11a then reduces to
$$
-\frac{c}{3}\frac{\partial f_1}{\partial r}+ 
\frac{\partial u}{\partial r} \frac{1}{3p^2}\frac{\partial (p^3 f_0)}{\partial p}=0
\eqno{(B2)}
$$
for energies above $cp_{max} $.
Integrating in space across the shock and in momentum from $p_{max}$ upwards gives
$$
j_{CR}=e \Delta u \frac {4\pi}{3} p^3 f_0(p_{max})
\eqno{(B3)}
$$
where $\Delta u$ is the change in velocity across the shock. 
Since $\Delta u = 3u_s/4$ for a strong shock
$$
j_{CR}=e\pi u_s p^3 f_0(p_{max})
\eqno{(B4)}
$$
as in equation 2.


\vskip 0.3 cm
\noindent
{\bf Appendix C: The derivation of equations 12 and 13}

Here we show how equation 12 and its vector equivalent can be derived from equations 11.
For simplicity we omit terms representing advection at the fluid velocity and the effect of the magnetic field.  These can easily be inserted at the end of the derivation.  The difficult part is the replacement of the stress tensor $f_{ij}$ by the vector $g_i$.
The most transparent way of presenting the derivation is to write it out in 
terms of individual components in $x$, $y$ and $z$ directions.
In the following we present the derivation of the equation
for ${\partial f_x}/{\partial t}$.  The derivations of the equations
for ${\partial f_y}/{\partial t}$ and ${\partial f_z}/{\partial t}$
follow the same pattern.
In component form, the relevant equations 11 are
$$
\frac{\partial f_0}{\partial t}=
-\frac{c}{3}
\left (\frac{\partial f_x}{\partial x}
+\frac{\partial f_y}{\partial y}+\frac{\partial f_z}{\partial z}\right)
$$
$$
\frac{\partial f_x}{\partial t}=
-c\frac{\partial f_0}{\partial x}
-\frac{2c}{5}
\left(
\frac{\partial f_{xx}}{\partial x}
+\frac{\partial f_{xy}}{\partial y}
+\frac{\partial f_{xz}}{\partial z}
\right)
$$
$$
\frac{\partial f_{xx}}{\partial t}=
\frac{c}{3}
\left(
-2\frac{\partial f_{x}}{\partial x}
+\frac{\partial f_{y}}{\partial y}
+\frac{\partial f_{z}}{\partial z}
\right)
$$
$$
\frac{\partial f_{xy}}{\partial t}=
-\frac{c}{2}
\left(
\frac{\partial f_{x}}{\partial y}
+\frac{\partial f_{y}}{\partial x}
\right)
$$
$$
\frac{\partial f_{xz}}{\partial t}=
-\frac{c}{2}
\left(
\frac{\partial f_{x}}{\partial z}
+\frac{\partial f_{z}}{\partial x}
\right)
\eqno{(C1)}
$$
Eliminating the components of the stress tensor between these equations gives
$$
\frac{\partial ^2 f_x}{\partial t^2 }
+\frac{9c}{5}\frac{\partial }{\partial t}\frac{\partial f_0 }{\partial x}
=
\frac{c^2}{5}
\left (
\frac{\partial ^2 f_x}{\partial x^2}
+\frac{\partial ^2 f_x}{\partial y^2}
+\frac{\partial ^2 f_x}{\partial z^2}
\right )
$$
$$
-\frac{c^2}{5}
\frac{\partial}{\partial x}
\left (
\frac{\partial f_x}{\partial x}
+\frac{\partial f_y}{\partial y}
+\frac{\partial f_z}{\partial z}
\right )
\eqno{(C2)}
$$
which is the $x$ component of the vector equation
$$
\frac{\partial ^2 {\bf f}_1}{\partial t^2}
+\frac{9c}{5}\frac{\partial (\nabla f_0)}{\partial t}
=
-\frac{c^2}{5} \nabla \times (\nabla \times {\bf f}_1)
\eqno{(C3)}
$$
and the same derivation holds for the $y$ and $z$ components of the equation.
This equation is second order in time differential.
It can be separated into two first order equations:
$$
\frac{\partial  {\bf f}_1}{\partial t}=
-\frac{9c}{5} \nabla f_0
-\frac{c}{5} \nabla \times {\bf g}_1
$$
$$
\frac{\partial  {\bf g}_1}{\partial t}=
c\nabla \times {\bf f}_1
\eqno{(C4)}
$$
These equations become equations 12 and 13 with the addition of 
fluid advection, rotation of ${\bf f}_1$ by the magnetic field,
and damping of the stress tensor at a rate $\nu _B$ 
as discussed above equation 12 in section 4.

Note the analogy of ${\bf f_1}$ and ${\bf g_1}$ in the above equations
with ${\bf E}$ and  ${\bf B}$ in Maxwell's equations.
In both cases they support transverse waves.

\vskip 0.5 cm
\noindent
{\bf Appendix D: The approximate CR equations: propagating modes}

Here we demonstrate that the equations set out in section 4
describe the essential CR propagation modes for monenergetic CR.
For simplicity we neglect the damping term ($\nu_B=0$). 
Propagating CR solutions in a stationary background plasma can be found by 
setting ${\bf u}=0$, $\partial /\partial r_i \rightarrow ik_i$
and  $\partial /\partial t \rightarrow -i\omega$ in equation (13):
$$
\omega f_0=\frac {c}{3}k_l f_l
$$
$$
\omega f_i = \frac {9c}{5} k_i f_0 + \frac{c}{5} \epsilon _{ijk} k_{j} g_{k} -i \epsilon _{ijk}\Omega_j f_k
$$
$$
\omega g_m = -c  \epsilon _{mpq} k_p f_q
\eqno{(D1)}
$$
The resulting dispersion relation is
$$
\left ( \omega ^2 -\frac{3c^2k^2}{5}  \right)
\left\{
\left ( \omega ^2 -\frac{c^2k^2}{5}  \right)^2
-\omega^2 \Omega^2 \right \}
$$
$$
-
\frac{2k^2c^2}{5}\omega ^2 \Omega^2 \sin ^2 \theta=0
\eqno{(D2)}
$$
where $\theta$ is the angle between ${\bf k}$ and ${\bf B}$.
There are two independent modes in the absence of magnetic field
($\Omega=0$).
The mode propagating at $\sqrt{3/5}c$ represents the motion 
of freely propagating CR with ${\bf f}_1$ parallel to ${\bf k}$.
The mode propagating at $\sqrt{1/5}c$ is the transverse mode
representing the motions of CR with ${\bf f}_1$ perpendicular to ${\bf k}$.
The transverse mode propagates more slowly because CR velocities are
aligned preferentially away from the direction of propagation.

In the presence of a magnetic field a longitudinal mode still propagates
parallel to ${\bf B}$ ($\theta =0$)
at $\sqrt{3/5}c$ representing free 
propagation along field lines unaffected by the field.
The transverse mode is relatively unaffected by the magnetic field
at wavelengths shorter than a Larmor radius $kc \gg \Omega$.
At wavelengths longer than the Larmor radius the transverse mode
propagates more slowly as the transverse CR current rotates rapidly in the magnetic field.

When mode propagation is directed across the magnetic field
($\sin \theta =1$)
the wave frequency is given by
$$
\omega^2=\Omega ^2 \left\{
\frac{1}{2}
+\frac{2k^2 r_g^2}{5}
\pm\sqrt{\frac{1}{4} + \frac{k^4 r_g^4}{25}+\frac{2 k^2 r_g^2}{5}}
\right \}
\eqno{(D3)}
$$
where $r_g$ is the CR Larmor radius.
In the limit of wavelengths smaller than the Larmor radius,
the frequency converges to those derived for 
the longitudinal and transverse waves in zero magnetic field as expected.
At long wavelengths ($kr_g \ll 1$)
the frequency converges to the Larmor frequency,
representing CR rotation in the magnetic field,
without significant propagation.

This analysis of the dispersion relation 
indicates that equations (13) for
$f_0$, ${\bf f}_1$ and  ${\bf g}_1$
provide an adequate representation of CR propagation.

\vskip 0.3 cm
\noindent
{\bf Appendix E: The approximate CR equations: the NRH instability}

\begin{figure}
\includegraphics[angle=0,width=8.6cm]{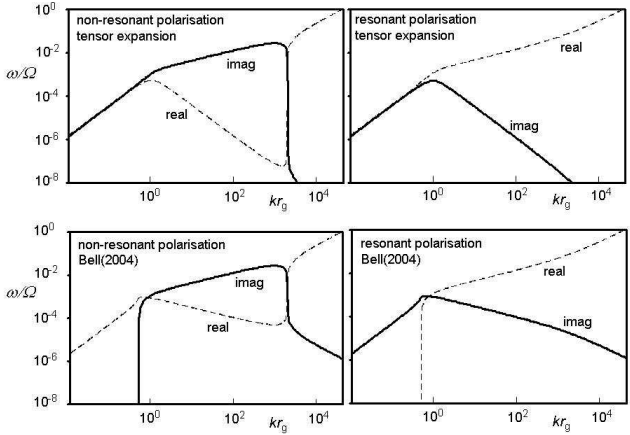}
\caption{
Dispersion relation for the resonant and non-resonant circular polarisations
as derived from the tensor expansion (top row)
compared with the dispersion relation derived by Bell (2004) 
for mono-energetic CR (bottom row).
The growth rates in units of the CR Larmor frequency
are given by full lines and the real frequencies by the 
dashed lines.  
Parameters relevant to the historical SNR are assumed.
}
\label{fig:figure5}
\end{figure}

We now investigate whether the approximate treatment of CR kinetics
in section 4
is adequate to model the NRH instability.
We derive the dispersion relation for the NRH instability 
in a simple case with the following assumptions.
The CR are mono-energetic with momentum $p$.
The zeroth order CR current ${\bf j}_0$,
the wavenumber ${\bf k}$ and the zeroth order magnetic field ${\bf B}_0$ are all parallel.
The background plasma is at rest to zeroth order, ${\bf u}_0=0$.
The first order perturbations to the magnetic field ${\bf B}_1$, CR current ${\bf j}_1$,
and plasma velocity ${\bf u}_1$ are all perpendicular to the zeroth order magnetic field
and the wavevector ${\bf k}$.
Since the modes are transverse
the plasma density is unperturbed to first order: $\rho _1 =0$.
The coupled linearised forms of equations (7), (13) and the Maxwell equation 
for $\partial {\bf B}/\partial t$ are then:
$$
\rho \frac {\partial {\bf u}_1}{\partial t}=-{\bf j}_0 \times {\bf B}_1-{\bf j}_1 \times {\bf B}_0
+\frac{1}{\mu _0}({\bf B}_0.\nabla){\bf B}_1
$$
$$
\frac {\partial {\bf B}_1}{\partial t}=({\bf B}_0.\nabla){\bf u}_1
$$
$$
\frac {\partial {\bf j}_1}{\partial t}=- \frac{c}{5}\nabla \times {\bf G}_1
-\frac {ec}{p} {\bf B}_0 \times {\bf j}_1-\frac {ec}{p} {\bf B}_1 \times {\bf j}_0
$$
$$
\frac {\partial {\bf G}_1}{\partial t}=c\nabla \times {\bf j}_1 -\nu _B {\bf G}_1
\eqno{(E1)}
$$
where ${\bf G_1}=(4\pi /3)ec\int p^2{\bf g_1}dp $.
For circular polarisation any first order perturbation  ${\bf \xi}_1$ 
(${\bf B}_1 $, $ {\bf j}_1 $, $ {\bf G}_1 $ or $ {\bf u}_1 $) satisfies
$$
\frac {\partial {\bf \xi}_1}{\partial t}=- \omega {\bf n}\times {\bf \xi}_1
\hskip 0.5cm
({\bf n}.\nabla ){\bf \xi}_1 = k {\bf n}\times {\bf \xi}_1
\eqno{(E2)}
$$
where ${\bf n}$ is a unit vector in the direction of ${\bf B}_0$,
giving
$$
-\rho \omega {\bf u}_1 = - j_0 {\bf B}_1 +B_0 {\bf j}_1 + \frac {kB_0}{\mu _0}{\bf B}_1
$$
$$
-\omega {\bf B}_1 = kB_0 {\bf u}_1 
$$
$$
-\omega {\bf j}_1 = - \frac {ck}{5} {\bf n}\times {\bf G}_1 -\frac{ecB_0}{p}{\bf j}_1 + \frac {ecj_0}{p} {\bf B}_1
$$
$$
-\omega {\bf G}_1 = ck {\bf n}\times {\bf j}_1 + \nu _B  {\bf n} \times {\bf G}_1
\eqno{(E3)}
$$
As discussed in section 4 we set the scattering frequency $\nu _B$
equal to the CR Larmor frequency.
For SNR conditions the NRH growth rate ($\sim \Gamma$)
is much less than the CR Larmor frequency, in which case
the dispersion relation simplifies to 
$$
\omega ^2  \approx  k^2 v_A^2
-  \sigma_\pm  \Gamma ^2 \left (1 \pm  \frac{ 5i }{k^2r_g^2} \right )^{-1}
\eqno{(E4)}
$$
where $v_A=\sqrt{B_0^2/\rho \mu _0}$ is the Alfven speed, 
$r_g$ is the CR Larmor radius in the magnetic field $B_0$
and 
$\Gamma =\sqrt { kB_0 j_0/\rho}$ is the NRH growth rate 
where it dominates in the range
$r_g^{-1} \ll k \ll \Gamma/v_A$.
$\sigma_\pm =\pm 1$ according to the sense of the circular polarisation
as determined by the sign of $k$.

For wavelengths much shorter than the Larmor radius ($k r_g \gg 1$)
$$
\omega ^2  = k^2 v_A^2
- \sigma_\pm  \Gamma ^2
\eqno{(E5)}
$$
which represents a purely growing instability for wavenumbers less than
$\Gamma /v_A$ in the appropriate polarisation $\sigma _\pm =1$.  
Tension in the magnetic field correctly damps waves with
wavenumbers greater than $\Gamma /v_A$ (Bell 2004).
The truncated tensor analysis is correct on scales smaller than the Larmor radius
because CR trajectories are then relatively unaffected by perturbations in the magnetic field. 
This is the important regime in which the rapidly growing NRH instability amplifies
the magnetic field.

At wavelengths longer than the Larmor radius ($kr_g\ll 1$)
the approximate tensor expansion gives
$$
\omega ^2  = k^2 v_A^2 \pm i   \frac{k^2 r_g^2}{5} \Gamma ^2
\eqno{(E6)}
$$
In comparison the correct dispersion relation in this limit is
$\omega ^2  = k^2 v_A^2 - {k^2 r_g^2} \Gamma ^2/5$ for the resonant instability
and $\omega ^2  = k^2 v_A^2 + {k^2 r_g^2} \Gamma ^2/5$ for the non-resonant instability
which in fact is stable for mono-energetic CR in this limit.
The Alfven term $k^2v_A^2$ is negligible at long wavelengths
so the tensor expansion gives a growing mode with
$$
 \omega =\frac{i\pm 1}{\sqrt{10}} (kr_g) \Gamma
 \eqno{(E7)}
$$
in both resonant and non-resonant polarisations.

Figure 5 compares the approximate tensor dispersion relation (top row)
with the correct dispersion relations of Bell (2004) (bottom row)
for both the resonant instability ($\sigma _\pm =-1$) that dominates for $kr_g<1$
and the rapidly growing non-resonant NRH instability ($\sigma _\pm =-1$) that dominates
for $kr_g>1$.
Crucially the tensor expansion accurately calculates the NRH growth rate in the range
$r_g^{-1}<k<\Gamma/v_A$. 
Outside this range the growth rate is too small for the instability to be effective.
The tensor expansion reproduces the growth of the resonant instability for $kr_g<1$
although it incorrectly produces weak growth of the non-resonant instability in this regime.
The crucial point for the simulation is that the tensor expansion
gives an accurate account of instability where the growth rate is large.


\vskip 0.3 cm
\noindent
{\bf Appendix F: Computational constraints}

The simulation makes heavy demands on computational resources
because it models spatial scales encompassing 
the wavelength of the fastest growing mode, the CR Larmor radius, 
and the free propagation
of CR ahead of the shock.
It is three-dimensional in configuration space and models the CR
distribution in momentum.
The spatial cell-size $\Delta x$ must be small enough to 
allow the NRH instability to grow from its initially small scale
($\Delta x < \pi/k_{max}$) and the timestep $\Delta t$
must be short enough to resolve CR crossing one computational cell
($\Delta t < \Delta x /c$).
The simulation must be run for a time $\tau$ at least 10 times the growth time
of the fastest growing mode $\tau =10\gamma_{max}^{-1}$
and the length of the computational grid $L_{||}$ in the direction 
parallel to
the shock normal must be large enough to allow the CR
to escape upstream: $L_{||}=10 c \gamma_{max}^{-1}$.
Provided the boundary conditions are periodic in the directions perpendicular
to the shock normal they can be much smaller than $L_{||}$
but they must be able to accommodate a CR Larmor radius:
$L_{\perp}=r_g$.
The number of computational operations is proportional to
$N_{comp}=\tau L_{||} L_\perp ^2/(\Delta t \Delta x ^3)$, which from
definitions and equations presented above, is of order
$$
N_{comp}\sim \frac {\eta^2 M_A^6}{4}
\eqno{(F1)}
$$

As discussed above the NRH instability grows strongly in a
wavenumber range between $r_g^{-1}$ and $k_{max}$.
Saturation due to magnetic tension occurs
on scalelengths comparable to the CR Larmor radius $r_{g,sat}$
when $B_{sat}/r_{g,sat}=\mu_0 j_{CR}$ as discussed in section 10.
The ratio of the saturated field  $B_{sat}$ to the initial field $B_0$
is $B_{sat}/B_0=(2r_g k_{max})^{1/2}$
where $k_{max}$ and $r_g$ are defined in the initial field $B_0$.
The equation for $k_{max}$ can be found in section 2, giving
$$
B_{sat}/B_0 \sim M_A \sqrt{\frac{\eta u_s}{c}}
\eqno{(F2)}
$$
where the acceleration efficiency $\eta$ is also defined in section 2.
The Alfven Mach number $M_A$ must be large
to allow significant amplification of the magnetic field ($B_{sat}\gg B_0$).
From equation (C1) large $M_A$ imposes a heavy demand on computational resources,
but from equation (C2) the cost can be minimised by making the shock velocity $u_s$ 
a large fraction of the speed of light.
We initiate the simulation with an unrealistically large magnetic field of 
$47 \mu{\rm G}$ to reduce the Alfven Mach number from its typical value of
$\sim 1000$ for shocks in young SNR.

\label{lastpage}

\end{document}